\documentclass[manuscript]{acmart}

\usepackage{enumitem}
\usepackage{color}
\usepackage{float}
\usepackage{stfloats}
\usepackage{soul}
\usepackage{tabularx}
\usepackage{multirow}
\usepackage{booktabs}
\usepackage{hyperref}
\usepackage{xcolor}
\usepackage{makecell}
\usepackage{graphicx}
\usepackage[most]{tcolorbox}
\usepackage{amsmath}
\usepackage{listings}
\usepackage{fvextra}
\usepackage{wrapfig}
\usepackage{subcaption}
\usepackage{array}
\usepackage{svg}
\usepackage{rotating}
\usepackage{threeparttable}

\newcommand{\sysname}{\textsc{DocuTeam}}
\newcommand{\treatment}{\sysname{}}
\newcommand{\control}{\texttt{Baseline}}
\newcommand{\stats}[7]{($\text{\treatment}=#1\pm#2$, $\text{\control}=#3\pm#4$, $#5=#6$, $p#7$)}
\newcommand{\myquote}[1]{\textit{``#1''}}
\definecolor{syscolor}{HTML}{3788F7}

  {\list{}{\leftmargin=0.2in\rightmargin=0in}\item[]\color{blockcolor}}%
  {\endlist}

\newcommand{\promptbox}[3]
{
    \begin{tcolorbox}[width=\linewidth, fontupper=\tiny]
        \textbf{#1} \\  % TITLE

        \hrule
        \bigskip

        \textbf{System Prompt} \\

        \begin{Verbatim}[breaklines, fontsize=\tiny]
#2
        \end{Verbatim}

        \hrule
        \bigskip

        \IfNoValueF{#3}{ % If argument #3 (User Prompt) is NOT empty
            \textbf{User Prompt} \\

            \begin{Verbatim}[breaklines, fontsize=\tiny]
#3
            \end{Verbatim}
        }
    \end{tcolorbox}
}

\definecolor{syscolor}{HTML}{3788F7}
\definecolor{blockcolor}{HTML}{555555}

\setcopyright{acmlicensed}
\copyrightyear{2018}
\acmYear{2018}
\acmDOI{XXXXXXX.XXXXXXX}

\acmConference[Conference acronym 'XX]{Make sure to enter the correct
  conference title from your rights confirmation email}{June 03--05,
  2018}{Woodstock, NY}

\acmISBN{978-1-4503-XXXX-X/2018/06}

\begin{document}

\title{\textsc{DocuTeam}: Mixed-Initiative Multi-Agent Discussions around Evolving Documents}

\author{Heechan Lee}
\email{hclee99@kaist.ac.kr}
\affiliation{%
  \institution{School of Computing, KAIST}
  \city{Daejeon}
  \country{Republic of Korea}
}

\author{Juhyeon Choi}
\email{wngus0223@naver.com}
\affiliation{%
  \institution{College of Liberal Studies, Seoul National University}
  \city{Seoul}
  \country{Republic of Korea}
}

\author{Tae Soo Kim}
\email{taesoo.kim@kaist.ac.kr}
\affiliation{%
  \institution{School of Computing, KAIST}
  \city{Daejeon}
  \country{Republic of Korea}
}

\author{Juho Kim}
\email{juhokim@kaist.ac.kr}
\affiliation{%
  \institution{School of Computing, KAIST}
  \city{Daejeon}
  \country{Republic of Korea}
}
\email{juho@skillbench.com}
\affiliation{
  \institution{SkillBench}
  \city{Santa Barbara, CA}
  \country{USA}
}

\author{Joseph Seering}
\email{seering@kaist.ac.kr}
\affiliation{%
  \institution{School of Computing, KAIST}
  \city{Daejeon}
  \country{Republic of Korea}
}

\renewcommand{\shortauthors}{Lee et al.}

\begin{abstract}
   In open-ended problem solving, collaborators often rely on discussion to surface concerns, challenge perspectives, and refine shared work as it evolves. While AI agents are increasingly used as discussion partners, existing multi-agent systems place a heavy burden on users to initiate and carefully orchestrate the discussions. We present \sysname{}, a mixed-initiative multi-agent discussion system in which both users and agents can initiate and steer conversations. Agents monitor document changes to proactively start and redirect discussions as the work evolves, while users can flexibly shape the conversation or adopt agent ideas. In a within-subjects study ($N=20$), participants using \sysname{} produced outcomes rated significantly more novel, relevant, and specific than with a baseline without any increase in cognitive load. Rather than using agents for one-off idea sourcing, participants engaged in an iterative refinement loop in which document changes prompted agent reactions, which led users to revisit and further develop their work.
\end{abstract}

% CCS


% KEYWORDS

% TEASER FIGURE
\begin{teaserfigure}
  \centering
  \includegraphics[width=1.00\textwidth]{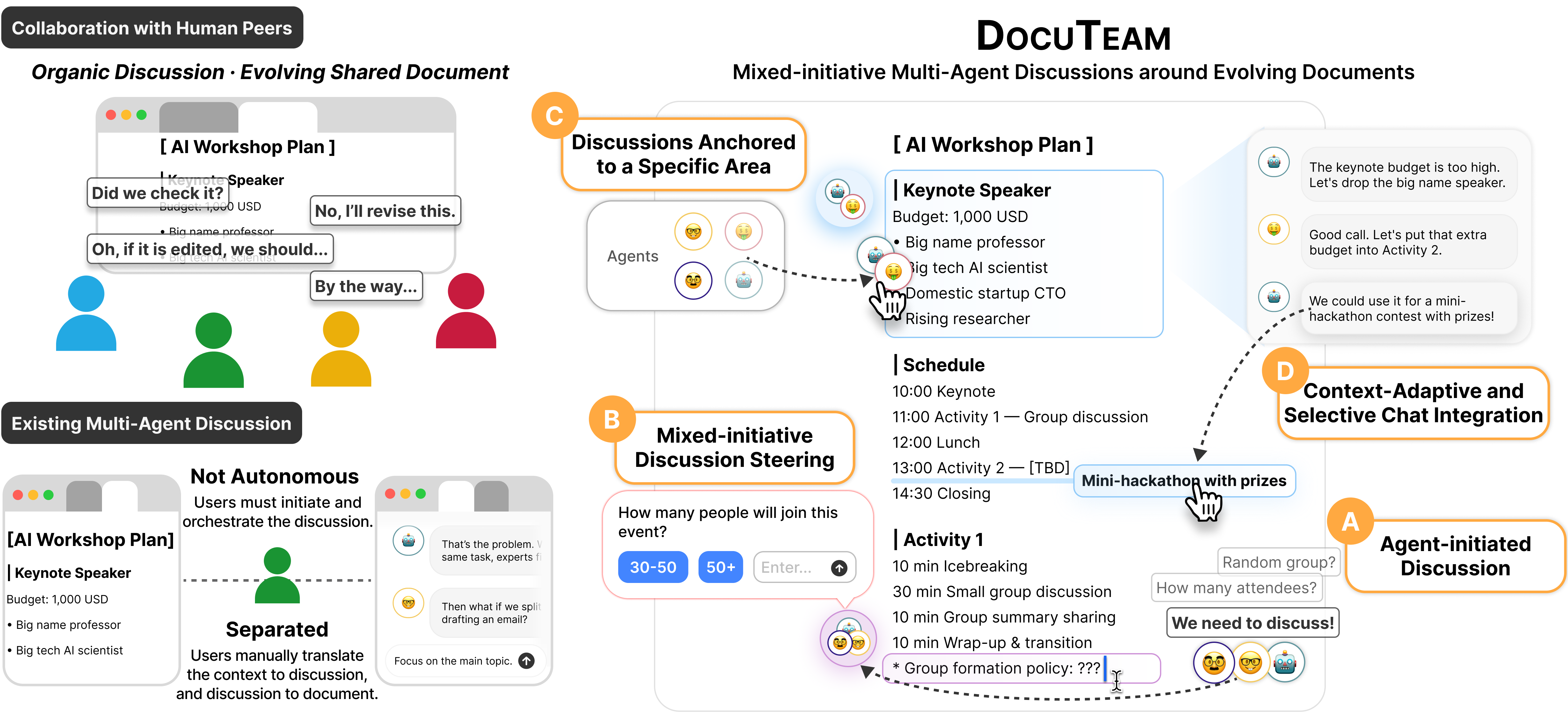}
  \caption{Illustration of a mixed-initiative multi-agent discussion interface.
 Whereas human teammates can organically build on and challenge one another's perspectives around shared work, existing multi-agent discussion systems often require users to initiate and orchestrate discussions separately from their working context. \sysname{} supports (A) organic agent-initiated discussions that proactively surface issues, (B) mixed-initiative discussion steering that allows both users and agents to guide the flow of conversation, (C) discussion sessions anchored to relevant document regions, and (D) selective incorporation of ideas from the discussion.}
    \Description{The figure contrasts human collaboration and existing multi-agent discussion systems on the left with DocuTeam on the right. At the top left, several human collaborators discuss an evolving shared document, illustrating discussion that emerges around ongoing work. At the bottom left, a conventional multi-agent interface separates the document from a side-panel chat and requires the user to initiate and sustain discussion. In DocuTeam (right), agent discussions are anchored directly to specific regions of an AI Workshop Plan document, demonstrating four core interactions. Agents are dragged onto a target document block and autonomously discuss its content beside that region, here debating whether to drop the big-name speaker and redirect the budget to Activity 2 (C). A discussion snippet, "Mini-hackathon with prizes," is applied into the TBD schedule slot and adapted to fit the surrounding context (D). Agents proactively initiate a discussion on an unresolved group formation policy line marked with question marks, surfacing questions like "Random group?" and "How many attendees?" without user prompting (A). When agents need user input to proceed, they surface a clarification question directly in the document, here asking how many people will join the event with selectable options (30--50, 50+) and a text input (B).}
  \label{fig:teaser}
\end{teaserfigure}

\maketitle

\section{Introduction}
In collaborative open-ended problem-solving, teammates often mix work with discussion, bringing different perspectives and expertise~\cite{hutchins1995cognition, hutchins2000distributed, page2019diversity, harvey2014creative, Arias2000TranscendingTI}.
Consider a student preparing a project report with their team: as the student revises the conclusion, they ask their team what they think about it.
One teammate mentions that the experiment results seem to strongly support the conclusion, while another urges more cautious interpretation due to the analysis's limitations. Another third member chimes in to note that a new figure in the results section conflicts with the explanation.
In the example, as the student works, they can initiate discussions about the work with their team members where they organically build on and challenge one another's perspectives~\cite{hutchins2000distributed, resnick1991perspectives, page2019diversity}, but, as the members also synchronously attend to the work~\cite{tang1991findings, gutwin2002descriptive, gergle2013using}, they can also proactively surface and engage in discussion points as the work develops~\cite{butchibabu2016implicit}---ultimately resulting in a stronger, more polished outcome.

With advancements in AI, recent studies have increasingly explored AI agents as a proxy for discussion partners, offering access to diverse perspectives and feedback when human collaborators are not available~\cite{gero2020mental, seeber2020machines, schelble2022let, zhang2021ideal}.
However, existing human-AI systems still fall short of enabling such team collaboration, where multiple AI agents autonomously start, drive, and redirect discussions as the work unfolds.
On one hand, prior work has proposed proactive agents~\cite{pu2025assistance, prasongpongchai2025talk, son2025clearfairy,choi2025proxona, lehmann2025collaborative, chen2026critiquecrew} that engage with users' ongoing work by providing contextually grounded feedback or performing operations (e.g., grammar-check) but rarely engage in and sustain agent-to-agent discussions, instead creating a ``hub-and-spoke'' model with the user at the center~\cite{liu2026from}.
On the other hand, systems where multiple agents can discuss with each other~\cite{liu2025perspectra, quan2025towards} rely on users to initiate, ground, and direct the discussions, instead of the agents proactively starting and directing discussions based on the user's ongoing work.
Taken together, these limitations leave underexplored the kind of mixed-initiative multi-agent discussion~\cite{horvitz1999principles} illustrated by the student example, where not only users can \textit{initiate} and \textit{develop} agent discussions but the agents themselves proactively engage in continued discussions grounded in the user's evolving work.

Realizing this form of mixed-initiative multi-agent discussions raises two main challenges.
First, as agents autonomously initiate and sustain multiple discussions alongside the user’s work, users must be able to follow and coordinate this activity without needing to continuously monitor it~\cite{pareek2026sensemaking, Schmbs2025FromCT}.
Second, agents should be able to carry exchanges forward without turn-by-turn prompting, but should be able to solicit user input when the discussion depends on unstated intentions or preferences.
To investigate these challenges and how to address them, we conducted an online design workshop with 15 participants who frequently use AI for open-ended tasks, such as writing, planning, and design, tasks commonly conducted by human teams using shared documents (e.g., text, sheet, canvases).
As a multi-agent discussion unfolds as an ongoing exchange and negotiation rather than an aggregation of isolated agent responses, participants wanted to be able to quickly follow the flow of discussion without disrupting their work, while retaining the ability to intervene directly or steer its direction indirectly by shaping its focus, participants, or emphasis.
Beyond grounding discussions in the work, participants also wanted to easily reflect discussions in their work by selectively bringing useful ideas or options from the agents back into their work.
From these findings, we derived four design implications for mixed-initiative multi-agent discussion systems: (1) support task-oriented discussion modes; (2) show discussions alongside the relevant parts of the work and summarize their current state; (3) enable both direct and indirect steering; and (4) support selective incorporation of ideas from agent discussions directly into the work.

Building on these design implications, we present \sysname{}, a mixed-initiative multi-agent discussion system that allows users and multiple AI agents to initiate, develop, and steer discussion---built on top of an existing document editor (i.e., Notion).
A discussion can begin from either side: users can initiate a discussion around a particular part of their work (Fig.~\ref{fig:teaser}C), while agents continuously attend to document changes and can proactively open a discussion when the evolving work raises an issue or opportunity (Fig.~\ref{fig:teaser}A).
Once initiated, agents can autonomously carry the discussion forward, building on and challenging one another through task-oriented discussion modes (i.e., Idea, Discussion, and Evaluation) without requiring users to orchestrate each turn.
\sysname{} keeps ongoing agent activity lightweight but legible by anchoring each discussion to the relevant document region and previewing its participants and current topic alongside the work (Fig.~\ref{fig:teaser}C).
As the discussion unfolds, users can step into a thread or change its participants to redirect the exchange, while agents can also ask clarification questions when further progress depends on the user’s intentions, preferred trade-offs, or unstated constraints (Fig.~\ref{fig:teaser}B).
When a discussion produces a useful idea, users can selectively move it back into the document, where it is adapted to the surrounding context (Fig.~\ref{fig:teaser}D), closing the loop between the evolving work and subsequent agent discussions.

To understand how \sysname{} affects users' task processes, outcome quality, and its impact on users' cognitive load in working with a team of agent collaborators, we conducted a within-subjects user study (N=20).
As a baseline, we compared our system against a setup that provides multi-agent discussion through a conventional side-panel interface, requiring users to initiate, ground, and steer agent discussions themselves.
We evaluated both systems on two open-ended event-planning tasks, where participants refined partially written plans under multiple constraints.
Independent human evaluators rated plans drafted with \sysname{} to be significantly more novel, relevant, and specific.
This improvement was accompanied by a shift in how participants collaborated with the multi-agent team during the task.
Compared to the baseline, participants spent less effort managing the discussion flow---sending 46\% fewer steering messages---while still making 39\% more document modifications than in the baseline condition, shifting from chat-centered idea sourcing toward document-centered iterative refinement.
Although overall cognitive load did not significantly differ, participants described a redistribution of effort: the mixed-initiative design introduced additional information that they had to process, but it also reduced the burden of orchestrating discussions and allowed more attention to remain on the document.
The document consequently became a shared medium for grounding interactions with the agents and iteratively incorporating diverse perspectives, helping explain the gains in novelty, relevance, and specificity.

Our findings point to the potential for mixed-initiative multi-agent systems to move beyond a group of agents that users must closely orchestrate. Although mixed-initiative multi-agent interaction is inherently difficult to design~\cite{pu2025assistance, chen2025maintaining, Schmbs2025FromCT, pareek2026sensemaking}, \sysname{} demonstrates that multi-agent discussion can become a viable form of team collaboration around evolving workflows.
However, realizing this model requires carefully balancing agent proactiveness with users' ability to understand and shape work processes.
We highlight directions for future research including investigating what balance of perspectives should optimally constitute an agent team and how much of an agent team's proactive activity should optimally be surfaced to the user.
\section{Related Work}
This section reviews relevant literature across three key areas: the foundational value of collaborative cognition through discussion, the current landscape of multi-agent collaboration, and interfaces for mixed-initiative AI agents.

\subsection{Collaborative Cognition through Discussions}
\label{sec:rw-first}
Complex, open-ended problems are often tackled by teams, where each member brings a distinct perspective and expertise~\cite{hutchins1995cognition, hutchins2000distributed, clark1991grounding, resnick1991perspectives, page2019diversity, harvey2014creative, Arias2000TranscendingTI}.
However, the benefits of diversity do not arise from simply assembling independent perspectives.
They depend on how teammates coordinate them through discussion: building on, challenging, and repairing one another's contributions to develop shared understanding~\cite{clark1986referring, clark1987collaborating}.
When this collaboration occurs around a shared artifact (e.g., document, workspace), effective teams will monitor the artifact to maintain awareness of each other's contributions in the artifact to proactively surface information or concerns~\cite{butchibabu2016implicit}, and rely on the artifact to coordinate actions and ground subsequent communication~\cite{tang1991findings, gutwin2002descriptive, gergle2013using}.
Together, these perspectives characterize collaborative problem solving as an ongoing process of coordinating diverse viewpoints through discussion grounded in shared work.

HCI and CSCW have long explored how computational systems can support the discussion structures for effective team collaboration.
Early groupware and electronic meeting systems introduced shared workspaces and explicit process structures to help groups externalize ideas, maintain awareness of one another's activity, and move between activities such as brainstorming, organizing, and evaluating alternatives~\cite{tatar1991cognoter, dourish1992awareness}.
Multiple lines of research have explored interfaces that help collaborators stay aware of ongoing discussions by structuring and summarizing group conversations~\cite{zhang2018making, chen2025meetmap, chandrasegaran2019talktraces}, as well as interfaces that connect communication to shared artifacts so that collaborators can recover the context, intentions and rationale behind evolving work~\cite{kim2021winder, xia2023crosstalk, li2025designmemo}.
More recent HCI systems have brought computational agents into group discussions, first as facilitators~\cite{kim2020botinthebunch, kim2021moderator, do2023err}, and increasingly as participants alongside human collaborators~\cite{seering2019beyond, zheng2023competent, yeo2026group}.
Building on this trajectory, we investigate how interfaces can make discussions among multiple AI participants a collaborative resource for users' ongoing document work, while supporting awareness of their activity and grounding their exchanges in the evolving artifact.

\subsection{Human--Multi-Agent Collaboration through Discussion}
AI systems have increasingly been studied not just as tools that execute user commands, but as collaborative partners that participate in joint tasks with humans~\cite{iftikhar2023human, bansal2019beyond, o2023human}.
Recent research has extended this paradigm from interaction with a single AI partner to collaboration with multiple specialized agents~\cite{gero2020mental, seeber2020machines, schelble2022let, zhang2021ideal}.
Multi-agent systems can distribute complex work across agents with complementary capabilities~\cite{hong2024metagpt, fourney2024magentic}, while emerging HCI research shows that users are beginning to form, interpret, and orchestrate teams of AI agents in creative and real-world work settings~\cite{lim2026understanding, pareek2026sensemaking, naik2025exploring}.
Within this broader landscape of human–multi-agent collaboration, one increasingly prominent interaction paradigm is multi-agent discussion~\cite{chan2024chateval, motger2026multi}.
In NLP research, structures where multiple agents debate or critique each other have shown promise in improving reasoning accuracy, sparking creative ideas, and enhancing evaluation quality through the comparison of diverse viewpoints~\cite{du2024improving, lu2024llm, haji2024improving, li2024mateval, Hu2024DebatetoWriteAP, Liang2023EncouragingDT}.

Building on this, HCI research has begun to show multi-agent discussions directly to users across domains such as unfamiliar decision-making~\cite{park2023choicemates}, collaborative ideation~\cite{quan2025towards}, creative support~\cite{choi2025proxona}, and scientific research~\cite{liu2025personaflow, liu2025perspectra}. This work shows that observing or steering agent-to-agent exchanges can help users discover alternatives, compare perspectives, and reflect more critically on their own decisions~\cite{he2025simupanel, zhang2024see, jiang2023communitybots, lippert2020multiple, Chiang2024EnhancingAG, hu2025dialoglab}.
However, existing multi-agent discussion systems are mostly reactive, requiring users to initiate with queries and directly manage the flow of discussions~\cite{fang2025llm, liu2025perspectra, choi2025proxona, quan2025towards}.
This creates the burden of sustaining the discussion on users, limiting the extent to which agents can function as an autonomous team that reacts to issues emerging in the work.

Furthermore, most multi-agent discussion systems remain detached from users' working artifacts, requiring users to manually bridge between agent conversations and their ongoing work~\cite{chen2025need, laban2024beyond}.
Recent work has begun to address this separation by integrating multiple AI agents directly into working artifacts: Lehmann et al.~\cite{lehmann2025collaborative} place user-configured agents in Google Docs, CritiqueCrew~\cite{chen2026critiquecrew} provides multi-perspective feedback around Figma components, and PaperDebugger~\cite{hou2026paperdebugger} embeds agents for reviewing, researching, and revising papers in Overleaf.
Yet, these systems primarily treat agents as individual consultants or functions that are invoked by user requests or predefined conditions, rather than as agents that autonomously develop discussions as issues emerge in the work.

Given the demonstrated benefits of multi-agent discussions for generating and surfacing diverse perspectives, our work explores how to make multiple agents autonomously initiate and sustain discussions and how to design them to be tightly integrated into users' working documents and workflows.

\begin{table*}[t]
\centering
\small
\setlength{\tabcolsep}{6pt}
\renewcommand{\arraystretch}{1.15}

\begin{tabularx}{\textwidth}{
    >{\raggedright\arraybackslash}p{0.17\textwidth}
    >{\raggedright\arraybackslash}X
    >{\raggedright\arraybackslash}X
    >{\raggedright\arraybackslash}X
}
\toprule
\textbf{Collaboration Systems}
&
\textbf{Agent-Initiated Interaction}
&
\textbf{Autonomous Discussion Development}
&
\textbf{Grounding in Evolving Work}
\\
\midrule

\textbf{Single Agent}
\cite{son2025clearfairy, pu2025assistance, prasongpongchai2025talk}
&
\textbf{Yes:} An agent proactively offers assistance and collaborates with the user.
&
\textbf{No:} No agent-to-agent discussion.
&
\textbf{Yes:} An agent accesses the document and observes users' behavior.
\\
\midrule

\textbf{Set of Multiple Agents}
\cite{choi2025proxona, chen2026critiquecrew, lehmann2025collaborative, fang2025llm}
&
\textbf{Limited:} Agents are invoked by user requests or predefined conditions.
&
\textbf{Limited:} Agents may exchange or synthesize perspectives, but do not autonomously initiate and develop open-ended discussions.
&
\textbf{Yes:} Each agent's contributions are grounded in the work artifact or context.
\\
\midrule

\textbf{Multi-Agent Discussion}
\cite{quan2025towards, liu2025perspectra}
&
\textbf{Limited:} Users initiate the discussion and manage when it occurs.
&
\textbf{Partial:} Agents debate, critique, and build on one another's perspectives, but require users' orchestration (e.g., via @mention or a ``Continue'' button).
&
\textbf{Limited:} Discussions are detached from the evolving work artifact.
\\
\midrule

\textbf{\sysname{}}
&
\textbf{Yes:} Agents proactively initiate discussions without requiring explicit user queries.
&
\textbf{Yes:} Agents autonomously develop and direct discussions with one another.
&
\textbf{Yes:} Discussions are grounded in the shared document as it evolves.
\\
\bottomrule
\end{tabularx}

\caption{
Comparison of interactions in Human-AI collaboration systems for document work.
Unlike prior systems, \sysname{} enables agents to proactively initiate and autonomously develop multi-agent discussions grounded in evolving work.
}
\Description{Table comparing four types of human-AI collaboration systems—single agents, sets of multiple agents, existing multi-agent discussion systems, and DocuTeam, across agent-initiated interaction, autonomous discussion development, and grounding in evolving work. Recent single-agent systems are proactive and grounded but lack agent-to-agent discussion; sets of multiple agents offer limited initiative and discussion development; existing multi-agent discussion systems support agent-to-agent discussion but typically require user initiation and are weakly grounded in evolving work. DocuTeam supports all three capabilities.}
\label{tab:rw}
\end{table*}

\subsection{Interfaces for Mixed-Initiative Human-AI Collaboration}
As described in \S~\ref{sec:rw-first}, human teammates do not simply wait to be instructed before every single contribution: they notice emerging needs, initiate new issues, challenge others' contributions, and sustain discussions.
HCI research has long examined how to design \textit{mixed-initiative} interactions~\cite{horvitz1999principles}, in which users and intelligent systems dynamically share control rather than acting as either fully autonomous actors or passive tools.
Foundational work further emphasizes that such interventions should be sensitive to uncertainty, user goals, timing, and the collaborative effort required to coordinate action~\cite{allen1999mixed, mcfarlane2002scope, erickson2003social}.

Recent HCI research has extended these principles to AI assistants that understand users' ongoing context and proactively intervene when assistance may be useful~\cite{kang2023synergi, pu2025ideasynth,  prasongpongchai2025talk, he2024ai, lee2024design}.
These studies have shown that grounding AI in local task context can make assistance more useful, and that proactive support can be effective when its timing and user control are carefully designed.
For example, systems have explored when programming assistants should proactively offer support~\cite{chen2025need, pu2025assistance}, how agents can work alongside users' ongoing activity through shared on-screen or real-time context~\cite{prasongpongchai2025talk, lee2025sensible}, and how in-situ agents can raise questions as users' artifacts evolve~\cite{son2025clearfairy}.

However, these principles have primarily been developed around interactions with a single agent.
Moving to multiple agents places additional coordination demands on users, who must make sense of multiple contributions while managing when and how agents participate.
Prior work has accordingly found that multi-agent interactions increase user confusion and turn-taking burden~\cite{Chaves2018SingleOM}, and more recent work identifies orchestration, conflict resolution, and sensemaking as central challenges for interactive multi-agent systems~\cite{Schmbs2025FromCT, pareek2026sensemaking, Zhang2025ConvoMapIV}.
For mixed-initiative multi-agent discussions, this challenge becomes particularly important because users must follow and steer discussions that unfold alongside their own ongoing work.
Existing interfaces largely address multi-agent complexity by giving users explicit control over agent participation and discussion structure~\cite{liu2025perspectra, quan2025towards}, leaving open how autonomous agent discussions can be integrated into users' task environment without creating additional coordination burdens.
To address this, our work investigates the interaction challenges of collaborating with a multi-agent discussion team and proposes an interface that enables agents to proactively initiate discussions grounded in the evolving document.
Table~\ref{tab:rw} summarizes these systems in terms of agent initiative, autonomous discussion development, and grounding in evolving work.
\section{Formative Study: Design Workshop}
Designing mixed-initiative multi-agent discussions requires agents that can autonomously initiate and develop discussions around evolving work, while allowing users to both stay focused on their work but also have awareness of the discussions and steer them when needed.
To understand how interfaces could support these competing demands, we conducted a design workshop to explore how such discussions should unfold alongside users’ work and how users should follow, intervene in, and act on them.

\subsection{Methods}
We conducted an online design workshop with 15 participants recruited from our institution’s online community who frequently (more than 3 times a week) used AI tools in open-ended problem solving with documents (e.g., writing with a text editor, planning with a spreadsheet, design with a digital canvas) and had experience designing user interfaces, as we expected participants to sketch possible design concepts during the workshop.
Participants joined 90-minute Zoom sessions in groups of three.
After a brief introduction to multi-agent discussion, each member of the group selected one of three familiar problem-solving scenarios---UI design, report writing, or house hunting, inspired by tasks previously used in \cite{chen2026critiquecrew, lehmann2025collaborative, pu2006increasing, park2023choicemates}.
They used a FigJam\footnote{\url{https://www.figma.com/figjam/}} board to brainstorm and sketch interface concepts.

The workshop focused on the following design questions:
\begin{itemize}
    \item DQ1: \textbf{What kinds of insights} do users expect from multi-agent discussion, and \textbf{when} during the task flow should insights surface?
    \item DQ2: \textbf{What interaction challenges} arise when multi-agent discussion is mixed-initiative and ongoing throughout document work, and \textbf{how} might interfaces address these challenges?
\end{itemize}

Two authors analyzed session recordings, transcripts, sticky notes, and interface sketches using thematic analysis to derive findings and design implications.
Detailed materials and procedures in Appendix~\ref{appendix:workshop_procedure}.
The study was approved by our institution’s IRB, and participants received 45,000 KRW (approximately 33 USD).

\subsection{Findings}
Participants envisioned multi-agent discussion as an ongoing flow that supports the task as the document and users' needs evolve throughout the work.
They also identified interaction challenges for making such discussions easy to follow, steer, and act on.

\subsubsection{What support do users expect from multi-agent discussion across the overall task flow?}
Participants expected multi-agent discussion to provide different kinds of support across the task flow, rather than serving a single fixed role throughout the work.
In particular, they expected three types of support from agents: first, (1) broaden possible options; next (2) examine trade-offs among alternatives; and finally (3) critique or validate emerging outputs.
These expectations resonate with classic work on argumentation-based multi-agent negotiation, which highlights how agents can surface and reconcile conflicting perspectives and coordinate toward acceptable solutions~\cite{Parsons1998AgentsTR}.

Rather than a smooth progression through these stages, participants expected all support types across all task stages---from initial goal setting and exploration to solution building and final review~\cite{Kuhlthau1991InsideTS, Suh2023SensecapeEM, Pirolli2005TheSP}---but with different emphasis at each stage.
During goal setting, participants imagined first establishing the criteria, priorities, and constraints that would guide the rest of the work, and they wanted agents to help refine these criteria and surface constraints that they didn't think of.
During exploration and formulation, they wanted agents to identify important information, fact-check, and probe underexamined areas.
During solution building, they wanted agents to compare alternatives, reveal trade-offs, and support convergence on reasonable options.
In final review, participants mentioned how it would be useful to provide agents with distinct personas that could then critique the document---e.g., giving agents personas such as clients, designers, or developers to identify issues in a UI design (P1 of Group 1, G1P1), or having each agent represent a specific evaluation criterion and provide comments aligned with that criterion (G4P2).
Representative examples and detailed support type--task stage mappings in Appendix~\ref{appendix:workshop_examples}.

\subsubsection{What interaction challenges arise from mixed-initiative multi-agent discussion?}
Beyond describing what agents should discuss, participants noted interaction challenges that would result from the flow of exchange and negotiation in mixed-initiative multi-agent discussion, contrasting with isolated responses from previous systems. Users needed to stay aware of how discussions were unfolding, intervene in their trajectories, and selectively act on the multiple ideas they produced.

\paragraph{Users wanted to stay aware of the flow of discussion without having to constantly monitor it.}
Across all groups, participants agreed that the discussion team should \myquote{keep talking in the background} (G3P3) or \myquote{provide suggestions or ideas steadily} (G5P1) even when the user was not leading the chat.
At the same time, they wanted to maintain awareness of how the current conversation was progressing.
G3P2 said, \myquote{The document shouldn't be blocked by agents ... but [when I want] I should be able to read quickly,} referring to a design in which agent discussions would not visually obscure the document or interrupt the user's primary work, while remaining easy to inspect on demand.
Participants proposed various interface designs for this purpose.
First, to give the user a sense that \textit{something is happening}, they proposed designs such as placing an indicator on the document region the discussion was currently focused on (G1P1), or displaying on the screen which agent was currently speaking (G3P1).
Second, they proposed an interaction that notifies the user when an important change occurs during the discussion.
For example, Group 5 came up with the idea of changing something like an ambient light's color on the screen when a change in situation occurred in the discussion.
Participants also emphasized the need to understand what the discussion was about without reading the full conversation.
All groups mentioned the need for a summary that allows key points to be quickly reviewed.
G1P1 said, \myquote{I don't think I need the full history ... I just want a summary with the key points,} and G4P3 wanted to \myquote{grasp the flow and relationships of the conversation rather than its specific content.}
Participants therefore wanted awareness of discussion to remain peripheral: enough to notice when a discussion was active or changing, understand its current direction, and inspect details only when needed.

\paragraph{Users wanted to steer the trajectory of discussion at different levels of engagement}
In all groups, participants wanted a function to steer the overall flow of the multi-agent conversation at a high level.
G1P3 emphasized that \myquote{if things are going strangely or the discussion is talking about something else, it is necessary for the user to intervene.}
Interestingly, participants wanted different levels of intervention.
The first was direct and strong steering, a method of intervening directly into the context of the conversation to express the user's opinion, such as by \myquote{jumping in} (G1P1) or \myquote{raising my hand} (G5P3).
The second was relatively indirect steering; in two groups, participants suggested giving weight to specific opinions through like/dislike functions (G2P2, G3P2).
Participants also considered designs such as adding a new agent to the ongoing conversation to represent the user's opinion, or excluding an agent that is saying strange or counterproductive things (G5P3).
G4P3 also proposed an interaction where agents or the current discussion is informed by pointing the mouse cursor over the parts of the document users want to emphasize.
These concepts were intended to allow the conversation to keep flowing while still allowing for indirect influence.
Participants thus treated steering not as turn-by-turn orchestration, but as the ability to intervene with different levels of effort --- from directly joining a thread to indirectly shaping its focus, participants, or relative emphasis.

\paragraph{Users wanted to selectively act on multiple ideas and versions emerging from discussion}
In 3 out of the 5 groups, participants proposed a function that reviews the various ideas emerging from the discussion and applies the ones they want to their work.
Participants saw discussion as generating multiple alternative ideas and evolving concepts for the document as it unfolded, allowing users to choose which ones to use rather than receiving a single final answer.
For example, G2P2 explained, \myquote{In the conversation, agents would come to propose various improvement ideas ... the user should be able to select the one they want among them.}
In addition, participants wanted to preview how these ideas would translate into their actual working document before deciding which ones to adopt.
G3P2 emphasized, \myquote{Rather than just reading the chat window, I want to see suggestions applied to the [document] I'm working on and choose based on that.}

\subsection{Design Implications}
Taken together, our findings suggest that making mixed-initiative multi-agent discussions useful for ongoing work requires more than simply placing agent discussions near the document.
Participants expected discussions to serve different purposes as their work evolved, while remaining easy to follow, steer, and selectively act on.
We therefore derive the following design implications.

\textbf{DI1. Support task-oriented discussion modes for expanding, deliberating, and evaluating.}
Participants expected discussion to serve different purposes across their work---from expanding possibilities to deliberating trade-offs and evaluating current documents---suggesting that systems should provide discussion modes aligned with these recurring task needs.

\textbf{DI2. Surface the focus, flow, and disagreements of ongoing discussions without requiring continuous monitoring.}
As discussions continue alongside document work, interfaces should help users see which parts of the document agents are discussing, quickly grasp how the discussion is developing and where agents agree or disagree through concise summaries, and bring emerging issues to the user's attention when needed.

\textbf{DI3. Support both direct and indirect steering of ongoing discussions.}
Participants wanted to intervene when discussions drifted from their intentions, but at varying levels of engagement; interfaces should therefore support both direct participation and lighter-weight ways of influencing the direction of discussion.

\textbf{DI4. Make ideas emerging from discussion easy to selectively incorporate into the document.}
Since discussion often produced multiple competing ideas and evolving versions, interfaces should make these alternatives easy to inspect and selectively incorporate into the user's ongoing work.
\section{\sysname{}: Mixed-initiative Multi-Agent Discussion on Documents}

\begin{figure*}[!t]
    \centering
    \includegraphics[width=1.00\textwidth]{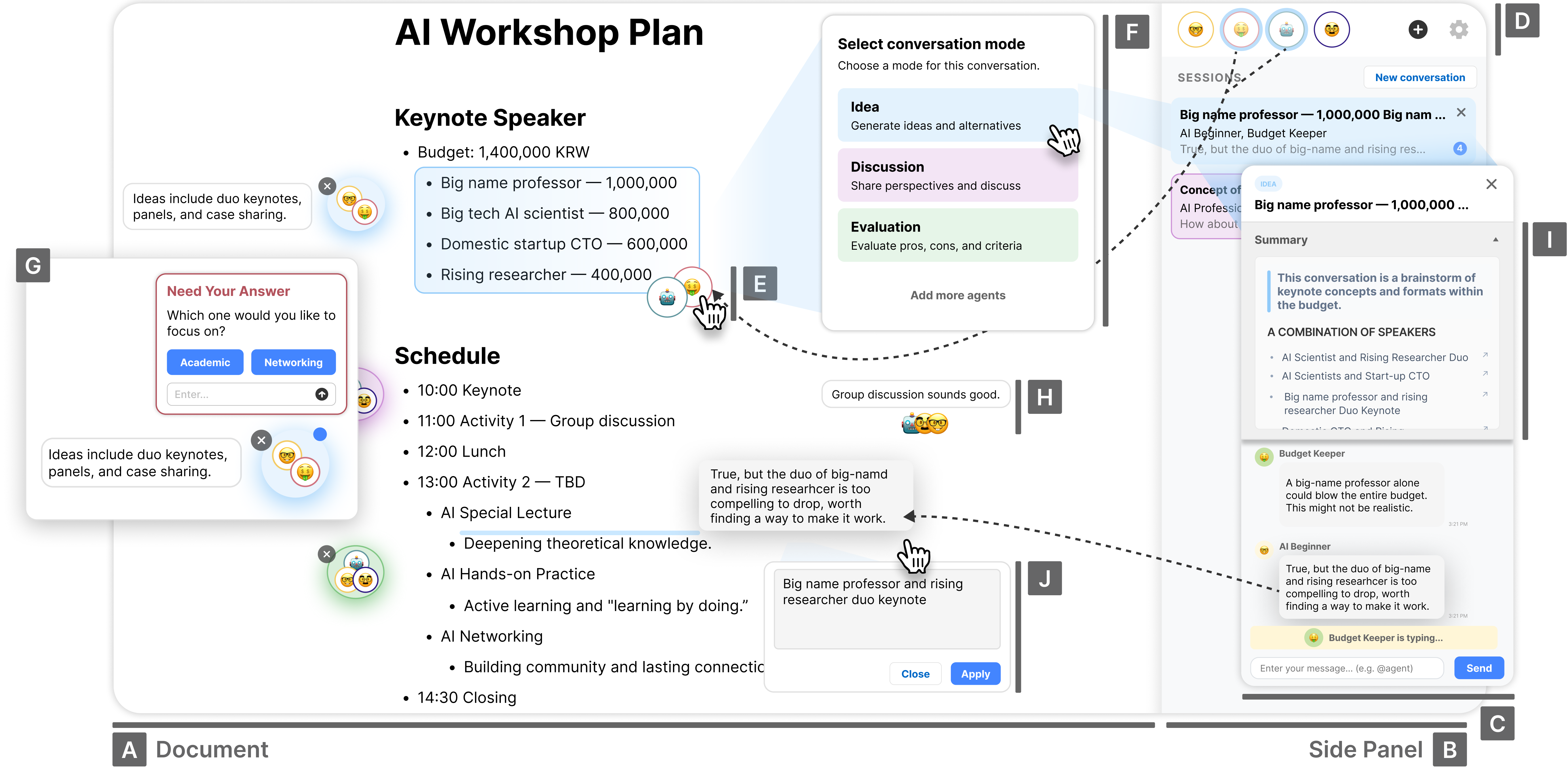}
    \caption{\sysname{} overview. \sysname{} consists of two main components: (A) a Document overlay view, where discussion sessions are anchored to specific regions of the document, and (B) a Side Panel, where users manage agents and inspect ongoing discussions. Users can initiate discussions by dragging agents onto the document and selecting a discussion mode (E, F). Discussions appear as anchored discussion bubbles beside the relevant document region (G) and can be opened in the side panel as chat threads with summaries (C, I). Agents proactively attend to edited blocks (H) and start discussions, while users can apply useful discussion content back into the document (J).}
    \Description{System overview of DocuTeam showing two main components. The left side displays an AI Workshop Plan document with discussion bubbles anchored beside specific blocks, showing agent avatars and message previews; one bubble has a red border prompting the user to choose a focus. The right side panel lets users manage agents, read ongoing chat threads, and view structured summaries of discussions. Users can start a discussion by dragging agents onto a document block and selecting a mode. A floating agent avatar beside an edited block indicates proactive monitoring. An Apply dialog lets users insert a selected discussion snippet directly back into the document.}
    \label{fig:system_overview}
\end{figure*}

Based on these design implications, we present \sysname{}, a mixed-initiative multi-agent discussion system that supports document-centered work in Notion\footnote{\url{https://notion.com/}}.
\sysname{} allows users to work directly in their document while multiple agents monitor the work and proactively discuss improvements and offer suggestions, allowing for user steering and input both directly and indirectly.
To align agent support with different stages of work, \sysname{} provides three task-oriented discussion modes---\textit{Idea}, \textit{Discussion}, and \textit{Evaluation} (DI1).
To provide users with lightweight awareness of ongoing discussions, rather than confining agent interactions to a separate chat interface, \sysname{} anchors agent discussions to specific document regions and presents summaries at varying levels of detail---from a one-line summary to a full discussion summary (DI2)---while still allowing users to view the full discussion if they so choose.
The system supports both user- and agent-initiated discussions and mixed-initiative steering, allowing both users and agents to shape ongoing conversations as needs emerge (DI3).
Users can also configure agent profiles and freely add or remove agents from ongoing discussion threads (DI3).
Finally, \sysname{} enables users to selectively integrate ideas from the discussion by dragging and dropping a specific chat message directly into the document (DI4).

\subsection{Interface Walkthrough}
\sysname{} consists of two main interface components.
The \textit{Document} view overlays \sysname{}'s interactive elements on top of a Notion page, which remains the primary document where users author and revise content; multi-agent discussion sessions are anchored to specific document blocks (Fig.~\ref{fig:system_overview}A).
The \textit{Side Panel} on the right serves as a hub for managing agents and for viewing all ongoing and closed conversation sessions (Fig.~\ref{fig:system_overview}B).
Through the side panel, users can read and inspect discussion threads, intervene by sending messages, and move useful discussion content back into the document.

\subsubsection{Agent Configuration}
To create multi-agent discussion teams that represent diverse perspectives, users can configure a team of agents tailored to their current task and goals.
Users can manage agents through a list displayed at the top of the side panel (Fig.~\ref{fig:system_overview}D).
For each agent, users specify a name, an avatar, and a brief description.
They can use these profiles to assign distinct roles or viewpoints---for example, a budget-conscious manager, an excitement-seeking event planner, or a conservative decision-maker.
Users can add a new agent via the \texttt{+} button, and can click an existing agent to edit or delete it.

\subsubsection{User-Initiated Discussions}
When users want to brainstorm, discuss or evaluate a specific part of the document, users can initiate a discussion by dragging one or more agents onto a target block in the document (Fig.~\ref{fig:system_overview}E).
They then select one of three discussion modes---\textit{Idea}, \textit{Discussion}, \textit{Evaluation}---and can optionally provide a short query to specify what they want the agents to focus on (Fig.~\ref{fig:system_overview}F).
Once the discussion starts, a \textit{Discussion Bubble} appears beside the associated block, visually anchoring the conversation to the relevant document region (Fig.~\ref{fig:system_overview}G).
The bubble shows the avatars of the participating agents, a one-line summary of the current discussion topic, and a blue dot when unread messages are available.
This bubble design helps users maintain awareness of where agents are focusing their attention and how the discussion is unfolding.
Users can open the full conversation by clicking the bubble or selecting the session from the side panel (Fig.~\ref{fig:system_overview}C).

\subsubsection{Agent-Initiated Discussions}
As users revise the document, \sysname{} guides the agents' attention by default toward recently edited blocks, as visualized through small floating agent avatars that appear to the right of the block the agents are currently reviewing (Fig.~\ref{fig:system_overview}H).
The agents assess whether ideation or further discussion is needed for the block and, if it is, they autonomously initiate a discussion.
The agents first determine a logical topic and select a discussion mode based on the recent changes; the underlying mechanism for this decision-making process is described in Section~\ref{sec:pipeline}.
The system then presents a discussion bubble beside the corresponding block, anchoring the resulting discussion to the relevant document region.
Although these bubbles function in the same way as in user-initiated discussions, they are visually distinguished through border styling.

\subsubsection{Reading and Using Discussions}
Users can read a discussion by clicking a discussion bubble or selecting it from the discussion list in the side panel (Fig.~\ref{fig:system_overview}C).
To help users quickly grasp long dialogues, the chat panel displays a mode-specific summary (Fig.~\ref{fig:system_overview}I).
The interface for each mode is detailed in Fig.~\ref{fig:appendix_summary}.
The bottom of the panel indicates which agent is preparing to speak next, or confirms that the conversation has concluded.
When users identify useful content while reading, they can drag-and-drop individual discussion snippets or summary items directly into the document, which are automatically adapted to match the surrounding content and writing style (Fig.~\ref{fig:system_overview}J).

\subsubsection{Mixed-Initiative Steering}
Rather than passively consuming agent-generated discussions, users can intervene at any point to redirect the conversation.
If the discussion is not unfolding as intended, users can directly send a message into the chat thread, introduce other agents that were not part of the original session, or kick specific agents from the discussion.
Conversely, when agents determine that user input is needed to resolve ambiguity or clarify priorities, they surface a question through a discussion bubble with a red border (Fig.~\ref{fig:system_overview}G).
Users can respond by typing a reply or selecting from a set of suggested response options.

\subsection{Discussion Pipeline}
Providing users with natural and insightful multi-agent discussions requires carefully designing both how diverse perspectives interact to produce productive exchanges and how the agents recognize appropriate moments and topics to initiate discussions from the document.
We present the design of \sysname{}'s discussion engine and its mechanism for deciding when to initiate discussions.

\subsubsection{Discussion Engine}
Inspired by prior work on multi-agent conversation design~\cite{Park2023GenerativeAI, liu2025perspectra, quan2025towards}, we designed a discussion engine to sustain meaningful exchanges across diverse perspectives while grounding the discussion in user and document context.
The engine autonomously guides the conversation toward productive insights while avoiding premature convergence across perspectives.
Each agent consists of two components: a \textit{profile}, which defines its role or viewpoint, and a \textit{long-term memory}, which stores persistent information (e.g., user preferences, prior decisions, and takeaways from earlier discussions).
The discussion engine operates in an iterative loop:

\begin{itemize}
    \item \textbf{Intent Extraction}: When a user initiates a discussion, the system prompts an LLM to infer the user's underlying intent and establishes a concrete goal for the discussion. This inference uses as context the current document content, task description, the focused block (where in the document the user dragged-and-dropped the agents), the selected discussion mode, and the user's query.
    \item \textbf{Decide Next Action}: A background moderator agent orchestrates the discussion flow. By evaluating the conversation history, the established goal, and the live document context, the moderator either (1) designates the most appropriate agent to speak next, (2) asks users to clarify information or preference between options, or (3) concludes that the goal has been sufficiently met.
    \item \textbf{Generate Utterance}: The designated agent generates its response based on its profile, the discussion mode and goal, the moderator's directive, the focused block, and the discussion history. Drawing from \textit{Perspectra}'s agent deliberation framework~\cite{liu2025perspectra}, the agent performs one of five actions: Issue, Claim, Support, Rebut, or Question.
    \item \textbf{Update}: After each turn, all participating agents update their long-term memories to integrate key takeaways from user messages, and newly formed consensus. Concurrently, an updated summary of the discussion is generated and streamed to the front-end. The update process runs asynchronously to minimize latency.
\end{itemize}

When a user sends a message, whether to a closed discussion or an ongoing one, the system immediately returns to the \textit{Decide Next Action} phase: reactivating closed discussions with a new goal, or overriding the current loop using the user's latest input.

\subsubsection{Agent-Initiated Discussions}
\label{sec:pipeline}
Inspired by Liu et al.'s guidance for proactive conversational agents~\cite{Liu2024ProactiveCA}, \sysname{} employs an automated discussion triggering mechanism.
The system monitors typing activity, applying a 2-second debounce to batch modifications and filter out trivial edits like typos.
Then, it analyzes the updated block and surrounding content, selects at least three relevant agents, and then these agents independently evaluate the necessity for a discussion---providing a score (0 to 1) for each of the three modes (i.e., Idea, Discussion, and Evaluation).
If the average score for any mode exceeds a predefined threshold of 0.75, which was determined through multiple pilot studies, the system automatically initiates a discussion.

\subsection{Implementation Details}
\sysname{} is implemented as a Chrome browser extension using ReactJS, CSS, and JavaScript.
We developed the extension to operate directly within Notion, chosen for its Markdown-based editing environment and the accessibility of its DOM, which allows for the reliable extraction of internal blocks and content.
The back-end was developed using Python and FastAPI.
Considering operational cost and latency within the LLM pipeline, we adopted \texttt{gpt-5-mini} for utterance generation, while using \texttt{gpt-4o-mini} for all other LLM components through OpenAI API\footnote{\url{https://platform.openai.com/}}.
Prompts are available in Appendix~\ref{appendix:prompts}.
\section{User Study}

To understand how users' open-ended problem solving processes are shaped by collaboration with a mixed-initiative multi-agent discussion team, we conducted a within-subjects study.
We compared \sysname{} to a baseline that provides the same multi-agent discussions but doesn't have the in-situ interactions and agent-initiated discussions.
The study was approved by our institution’s IRB.
Through this study, we focus on answering the following research questions:
\begin{itemize}
    \item \textbf{RQ1. (Outcome)} How do mixed-initiative multi-agent discussions impact novelty, workability, relevance, and specificity of the outcomes of open-ended problem solving?
    \item \textbf{RQ2. (Process)} How do mixed-initiative multi-agent discussions change users’ patterns of interactions with agents during the task?
    \item \textbf{RQ3. (Cognitive workload)} How does working with mixed-initiative multi-agent discussion teams affect the cognitive and coordination burdens placed on users during the task?
\end{itemize}

\subsection{Study Design}

\subsubsection{Participants}

We recruited 20 participants through our institute's online communities who regularly use LLM services to discuss or solve complex problems and had at least one prior experience planning an event.
Participants were compensated 50,000 KRW (approximately 37 USD) for the 2-hour study.

\subsubsection{Condition}

We compared the full \sysname{} system against a \texttt{Baseline} condition representing a conventional user-initiated multi-agent discussion interface.
Since prior work has already shown that multi-agent discussion can offer advantages over single-agent settings in similar tasks~\cite{liu2025perspectra, quan2025towards}, we adopted a multi-agent baseline to provide a stronger comparison rather than comparing \sysname{} against a single agent.
The \texttt{Baseline} condition was a version of \sysname{} without agent-initiated discussion, steering, and in-situ interaction in documents---functionally a user-initiative system rather than a mixed-initiative system.
Instead of using drag-and-drop to initiate discussions anchored to specific areas of the document, users in the \texttt{Baseline} condition initiated chat threads grounded in the whole document from a separate panel.
To ensure a fair comparison, both conditions were powered by the identical discussion engine and provided auto-generated summaries of discussions using the same pipeline.

\subsubsection{Tasks}

We designed two open-ended event-planning tasks: \textit{Academic Workshop Planning} and \textit{Joint Sports Day Planning}.
We selected event planning because it provides an accessible, real-world setting in which participants can consider multiple constraints and stakeholder perspectives while iteratively refining documented outcomes (e.g., timetables, event programs)~\cite{kim2013cobi}.
In addition, event planning shares several key characteristics with tasks from the formative study. Similar to house hunting, it involves comparing multiple options. Like UI design, it requires reconciling various constraints into a coherent plan. Finally, it involves documenting decisions and reflections, like report writing.
In both tasks, participants were given a document with a partially written plan and reference materials (e.g., constraints).
We provided an initial plan rather than starting from scratch to help participants quickly establish task context and to ensure that the task included opportunities for different forms of discussion, such as generating ideas and deliberating among constrained alternatives.
Participants were tasked with improving the existing plans, while also addressing the discussion points.
Participants were provided with a few predefined agents, and they could add additional agents representing other perspectives or requirements if needed.
Full study materials are provided in Appendix~\ref{appendix:user_study_materials}.

\subsubsection{Procedure}

The study procedure began with obtaining informed consent from each participant, and each study session lasted approximately 2 hours. 
After a research briefing (5 min), participants completed two task blocks, one for each condition.
The order of conditions and tasks was counterbalanced.
Each task block lasted 45 minutes and consisted of four stages: a system tutorial (5 min), a task explanation (5 min), the main event-planning task (30 min), and a post-survey (5 min).
Before starting each task, participants were given time to review the current draft and the provided reference materials.
After completing both task blocks, participants participated in a semi-structured interview (about 20 min).
The interview focused on how they used the agent discussions during the task, how the interface affected their planning process, and how they perceived the usefulness and limitations of the two conditions.

\subsubsection{Measures}

To assess the quality differences in the generated proposals across the two conditions, we conducted a blind expert evaluation.
We recruited two experienced external evaluators with over two years of experience in planning university festivals and on-campus workshops, compensating them with 100,000 KRW (approximately 70 USD) each.
Raters evaluated the 40 final plans produced by the 20 participants, rating them on a 5-point Likert scale across four dimensions: Novelty, Workability, Relevance, and Specificity, drawn from \cite{Dean2006IdentifyingQN}.
For quantitative subjective feedback, we analyzed post-survey responses where participants used 7-point Likert scales to rate the extent to which the system expanded or deepened their thinking during the planning process, their confidence and satisfaction with the final outcome, and their perception of the dialogues with the AI agents.
Participants also completed a NASA-TLX questionnaire~\cite{hart1988development}, excluding `Physical Demand', to report their perceived workload.
Likert scale responses were analyzed using the Wilcoxon signed-rank test.
We also recorded participants' interaction logs, including: use of each dialogue mode, dialogue timestamps, messages which participants sent to intervene discussions, and each modification to a document.
During the study, we counted each time a participant incorporated an idea or suggestion from the discussions into their document, and we confirmed each instance with the participant after the session.
For these interaction-based measures, we ran Shapiro-Wilk tests to determine if the data was parametric, and then adopted a paired t-test (if parametric) or a Wilcoxon signed-rank test (if non-parametric).
For qualitative data, we coded the participants' comments from the semi-structured interviews through thematic analysis.
The full survey questions are provided in Appendix~\ref{appendix:user_study_questions}.
\subsection{Results}

In this section, we report how \sysname{} affected the quality of participants' final outcomes (RQ1), their interaction patterns with mixed-initiative multi-agent discussions (RQ2), and their relative cognitive burdens (RQ3).

\begin{figure*}[!t]
    \centering
    \includegraphics[width=0.95\textwidth]{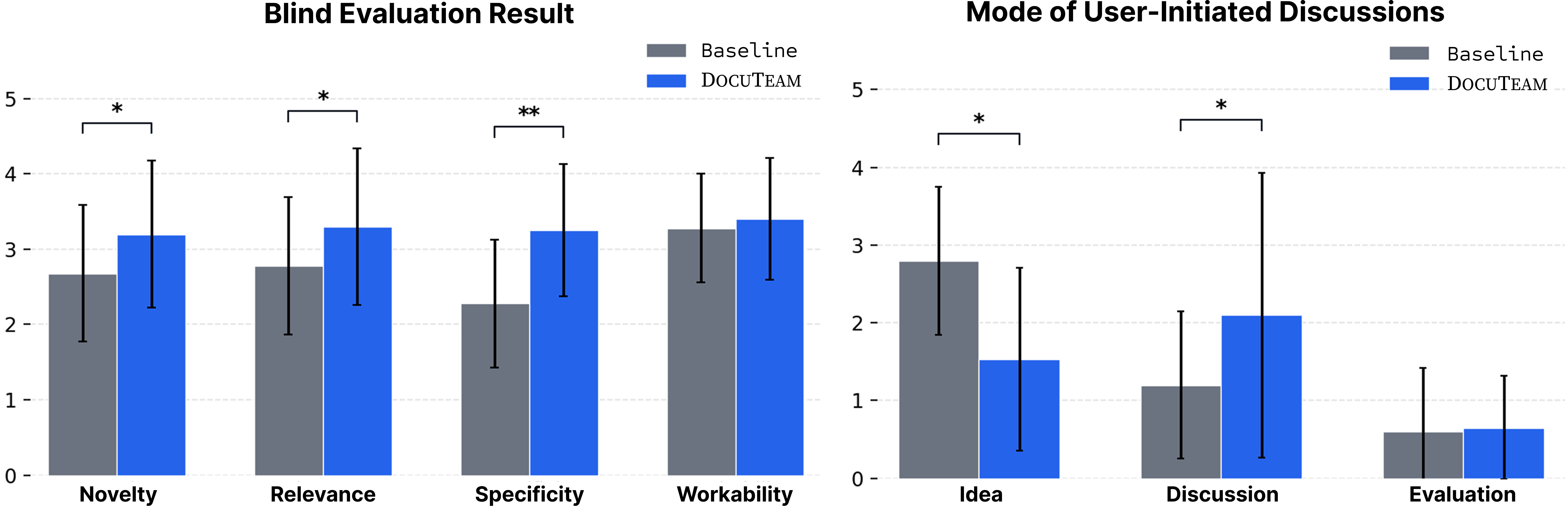}
    \caption{Blind evaluations showed that the \treatment{} condition produced more novel, relevant, and specific outcomes than the \control{} condition. The mode of user-initiated discussions shifted from Idea to Discussion. (*:p<.05, **:p<.01, error bars indicate one standard deviation)}
    \Description{Two grouped bar charts comparing Embedded (blue) and Detached (gray) conditions, with error bars indicating one standard deviation. The left chart shows blind evaluation scores (0--5 scale) across four quality dimensions. Embedded scores higher than Detached on Novelty (3.20 vs. 2.78, p<.05), Relevance (3.30 vs. 2.78, p<.05), and Specificity (3.25 vs. 2.28, p<.01), while Workability shows no significant difference (3.4 vs. 3.28). The right chart shows the frequency of user-initiated discussion modes (0--5 scale). Embedded participants used the Idea mode less than Detached (1.53 vs. 2.8, p<.05), used Discussion mode more (2.1 vs. 1.2, p<.05), and Evaluation mode was similarly low in both conditions (approximately 0.6).}
    \label{fig:result_blind_and_mode}
\end{figure*}

\subsubsection{(RQ1) More novel, relevant, and specific outcomes}
\label{sec:result-rq1}
In condition-blind evaluations, the final plans produced with the \treatment{} condition were rated significantly higher in Novelty \stats{3.20}{0.98}{2.68}{0.91}{W}{35}{=0.047}, Relevance \stats{3.30}{1.04}{2.78}{0.91}{W}{40.5}{=0.048}, and Specificity \stats{3.25}{0.88}{2.28}{0.85}{W}{13}{=0.001}, although with no significant differences in Workability \stats{3.40}{0.81}{3.28}{0.72}{W}{51.5}{=0.623}.
This suggests that \sysname{} supported not only the generation of more original ideas, but also the development of plans that better reflected task requirements and were articulated in greater detail.

\textbf{Agent-initiated discussions and in-situ interaction contributed to more novel, relevant, and specific outcomes.}
The interviews suggested two distinct mechanisms behind these gains.
First, agent-initiative discussions proactively exposed users to perspectives and suggestions they would not have generated on their own, contributing to higher novelty.
As P6 noted, \myquote{I could add points I hadn't thought of—it felt like I wasn't confined [to my own thoughts].}
For relevance, proactive exchanges among agents surfaced counterarguments, risks, and overlooked constraints before users explicitly raised them, helping participants better align their evolving drafts with task requirements.
For example, P14 mentioned, \myquote{I was considering what potential issues might arise, and they were proactively discussing safety. Seeing that made me realize I needed to prepare for it.}
The \treatment{} condition supported specificity as discussions were anchored to specific blocks in the document, which enabled \myquote{specialized discussions that were divided by section} (P20), allowing participants to iteratively refine what they had already written.
External evaluators explained that the generally similar workability scores stem from incompleteness: regardless of condition, participants struggled to develop complete plans that could be used for real events due to the study's time constraints.

\textbf{While outcomes improved, competing perspectives made participants feel uncertain of the outcome quality.}
Although \treatment{} led to higher-rated outcomes, these improvements did not translate into higher self-reported confidence \stats{4.35}{1.63}{4.40}{1.47}{W}{50.5}{=0.896} or satisfaction \stats{5.15}{1.50}{5.15}{1.23}{W}{44.5}{=0.943}.
Participants sometimes felt less certain about the quality of their work because the agents not only introduced counterarguments and alternative perspectives that \myquote{contradicted [their] ideas} (P13), but also broadened the range of issues to consider, making it \myquote{harder to organize everything} (P17).
This contrast suggests that \treatment{} may have improved final outcomes by encouraging more critical reflection, which in turn hindered users’ subjective confidence.

\subsubsection{(RQ2) From chat-centered ideation to document-mediated discussion and iterative refinement}
\label{sec:result-rq2}
\textbf{The document became a medium through which users and agents collaborate with one another.}
\treatment{} changed how participants engage with multi-agent discussions in their work.
In \treatment{}, participants initiated Discussion-mode conversations significantly more often \stats{2.1}{1.83}{1.2}{0.95}{t}{-2.35}{=0.030}.
In contrast, in \control{}, participants relied more on broad ideation, with user-initiated Idea-mode conversations occurring significantly more often \stats{1.53}{1.18}{2.8}{0.95}{t}{2.59}{=0.018}.
P6 explained, \myquote{Because I could place agents on a specific part, I ended up using Discussion-mode more. Ideation starts from zero, whereas discussion builds on what is already written.}
This suggests that \sysname{} shifted the role of multi-agent discussion from external idea generation toward developing and refining content already present in the document.
This pattern was further reinforced by the agents’ proactive and autonomous discussions on users’ ongoing task.
P13 described, \myquote{Once I put something into the document, the agents would discuss it on their own, which made me look at it again. ... In \control{}, I would not get that additional feedback unless I asked for it,} and P19 agreed that agent-initiated discussions \myquote{made me think one more time}.
Importantly, this implied that writing itself could become part of the interaction with the agents.
Some participants wrote in anticipation of such reactions; as P2 described, agent initiative \myquote{made me keep typing more as much as possible to see agent reactions.}
In this way, the document served not only as the object being edited, but also as a medium through which users and agents continually responded to one another.
Together, these patterns suggest that \treatment{} shifted multi-agent discussion from one-shot idea generation toward an iterative refinement loop around the evolving document, which explains the improved specificity observed in \S~\ref{sec:result-rq1}.

\begin{figure*}[t]
    \centering
    \includegraphics[width=0.80\textwidth]{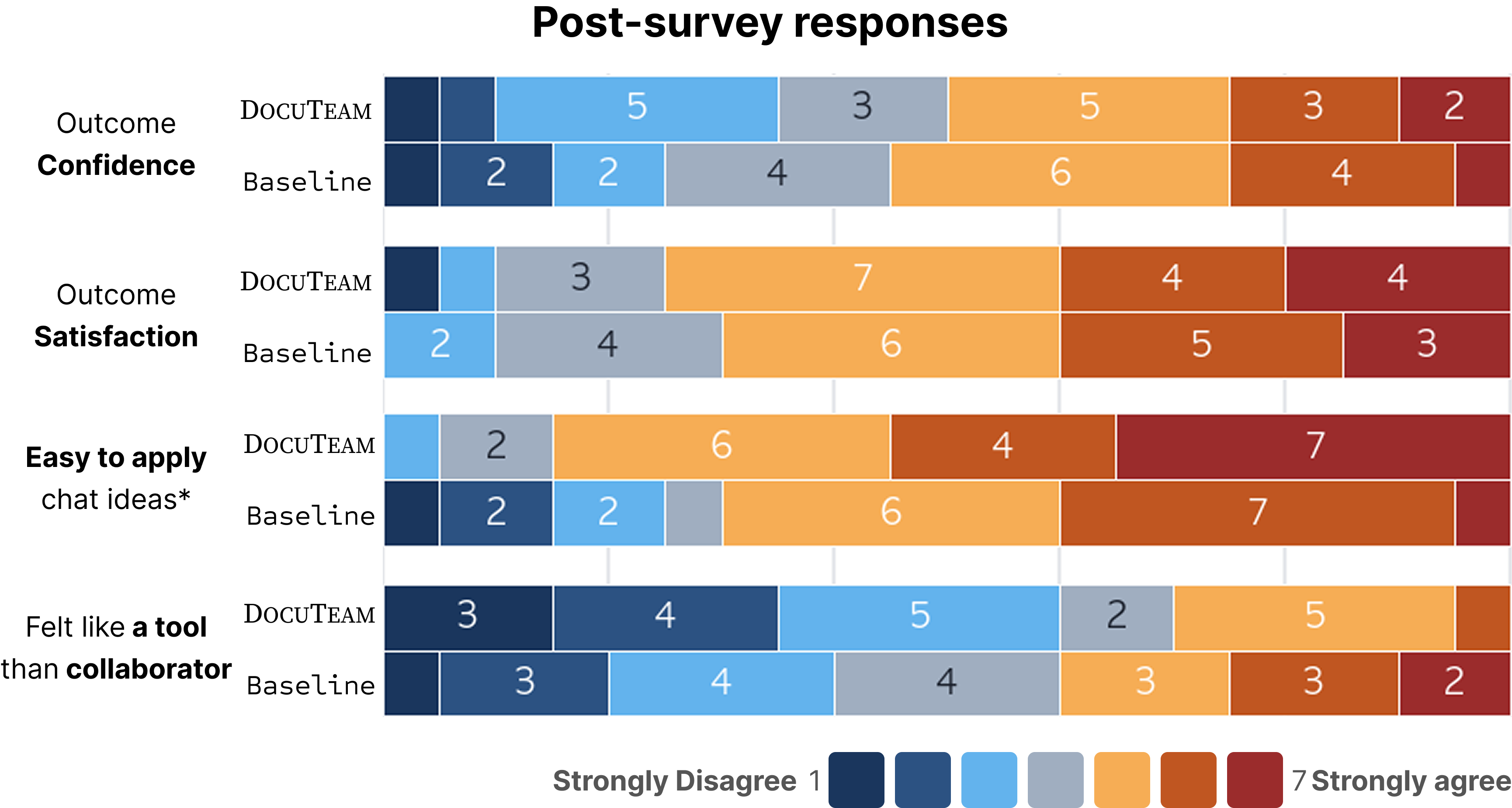}
    \caption{Post-survey responses on outcome and collaboration. Unlike the external evaluators' assessments, participants' confidence in and satisfaction with their own outcomes did not differ significantly between conditions. \treatment{} made discussion ideas easier to apply to the document. (*:p<.05)}
    \Description{Diverging stacked bar charts using a 7-point Likert scale from Strongly Disagree (dark navy) to Strongly Agree (dark red), comparing DocuTeam and baseline conditions. It shows six outcome and collaboration items. Outcome Confidence and Outcome Satisfaction show similar distributions across conditions, both skewing toward agreement. Felt like a tool than collaborator (reversed) are similarly distributed between conditions. Easy to apply chat ideas is the only significant item (p<.05), with DocuTeam scoring higher than baseline, indicating DocuTeam made it easier to apply discussion ideas to the document.}
    \label{fig:result_likert}
\end{figure*}

\textbf{This iterative refinement also made ideas from multi-agent discussions easier to incorporate into the draft.}
Because discussions in \treatment{} were grounded in specific parts of the document, participants could obtain more contextually relevant discussions and apply them more directly.
P11 said \myquote{Placing the agent in a specific part was the best... it was good to get context-specific help.}
As shown in Figure~\ref{fig:result_likert}, participants reported that it was significantly easier to apply AI-generated ideas \stats{5.7}{1.22}{4.6}{1.43}{W}{49}{=0.025}.
Consistent with this, they added or modified significantly more words in their final drafts \stats{448.7}{100.95}{322.95}{135.33}{t}{3.68}{<0.01} and applied more messages from discussions to the final document \stats{12.3}{3.57}{9}{3.36}{t}{4.71}{<0.01}.

Participants in the \treatment{} condition reported slightly lower scores on the survey question assessing whether the AI felt like a mere tool rather than a collaborator \stats{3.25}{1.55}{4.1}{1.74}{W}{19.0}{=0.060}, although this difference was not statistically significant.
Interviews help contextualize this: participants described the agents as feeling more \myquote{alive} (P15, P18) and sometimes even \myquote{cute} (P3, P5, P15, P16) in the \treatment{} condition, reflecting how their visible and proactive activity made them seem less like passive tools and more like companions working alongside them.
As P8 mentioned, \myquote{It gave me peace of mind because I can feel they were working on their own. Sometimes distracting, but psychologically reassuring.}

% Preamble
% \usepackage{booktabs}
% \usepackage{threeparttable}

\begin{table}[b]
\centering
\caption{Differences in task process measures between \control{} and \treatment{}.}
\Description{Table comparing three task process metrics between the treatment and control conditions. The treatment group sent fewer steering messages than the control group (2.70 vs. 5.00), suggesting less need for correction. However, the treatment group made more word-level draft changes (448.70 vs. 322.95) and applied more messages to the document (12.30 vs. 9.00), indicating more active document editing. All values are means with standard deviations. Differences in steering messages are significant at p < .05; draft word changes and messages applied to document are significant at p < .01.}
\label{tab:process_results}
\begin{threeparttable}
\begin{tabular}{lcc}
\toprule
Metric & \treatment{} & \control{} \\
\midrule
Steering msgs.*   & $2.70 \pm 0.77$   & $5.00 \pm 1.50$ \\
Draft word changes.** & $448.70 \pm 100.95$ & $322.95 \pm 135.33$ \\
Msgs applied to doc.**   & $12.30 \pm 3.57$  & $9.00 \pm 3.36$ \\
\bottomrule
\end{tabular}

\begin{tablenotes}[flushleft]
\footnotesize
\item Values are reported as $M \pm SD$. 
\item * $p < .05$, ** $p < .01$.
\end{tablenotes}
\end{threeparttable}
\end{table}

\subsubsection{(RQ3) Comparable overall workload despite different interaction demands}
Despite the additional proactive and concurrent interactions introduced in \treatment{}, participants did not report significantly higher overall cognitive demands.
NASA-TLX scores were similar in both conditions \stats{4.65}{0.59}{4.26}{0.89}{W}{47.5}{=0.169}, suggesting no significant change in perceived workload.
The interviews show that this null result reflected countervailing experiences: while \treatment{} introduced additional cognitive demands as participants had to attend to the agent discussions that would initiate throughout their document, it also reduced the effort required to direct these discussions and provided scaffolds to keep track of them.
Some participants found the amount and concurrency of agent activity somewhat overwhelming.
For example, P15 described, \myquote{There was a lot of information coming from multiple places, so it felt a little overwhelming.}
Participants also described an initial learning cost in adapting to the more active interface.
As P16 described, \myquote{At first it felt a little overwhelming, but once I became familiar with the system and how the agents provided ideas, it actually helped me become more immersed and expand my thinking.}

\textbf{Attention shifted from discussion steering toward document work itself.}
Participants sent significantly fewer messages to steer the discussion in the \treatment{} condition \stats{2.7}{0.77}{5}{1.50}{W}{31.0}{=0.013}.
While, in isolation, this result could have many possible explanations, participants' interviews made it clear that the proactive nature of the agents reduced their need to continuously steer the discussion.
Because the agents were already grounded in the evolving document, participants could focus on the document itself rather than continually restating and refining their intentions through chat.
As P16 explained, \myquote{In \treatment{}, the agents look at what I am writing on, so once I assign a specific block, I do not have to think about how to steer the conversation. In \control{}, I had to explain things through chat and keep directing it the way I wanted.}
% P13 also mentioned, \myquote{In \control{}, I had to focus on chat and type a lot, whereas in \treatment{}, the agents would attach themselves and start discussing on their own, so it felt like that took less effort.}
For some participants, this reduction in conversational management allowed more attention to remain on the document itself.
P10 described, \myquote{\treatment{} made me focus more on the document. I selected areas to ask about, and I ended up reading the document a lot. But, in \control{}, [I was] focusing more on chat to phrase my questions and explain my intentions through chat.}
This reduced the coordination burden of managing multi-agent discussion alongside writing, making the interaction feel more like a continuation of document work than a separate conversational task.

\begin{figure*}[t]
    \centering
    \includegraphics[width=0.70\textwidth]{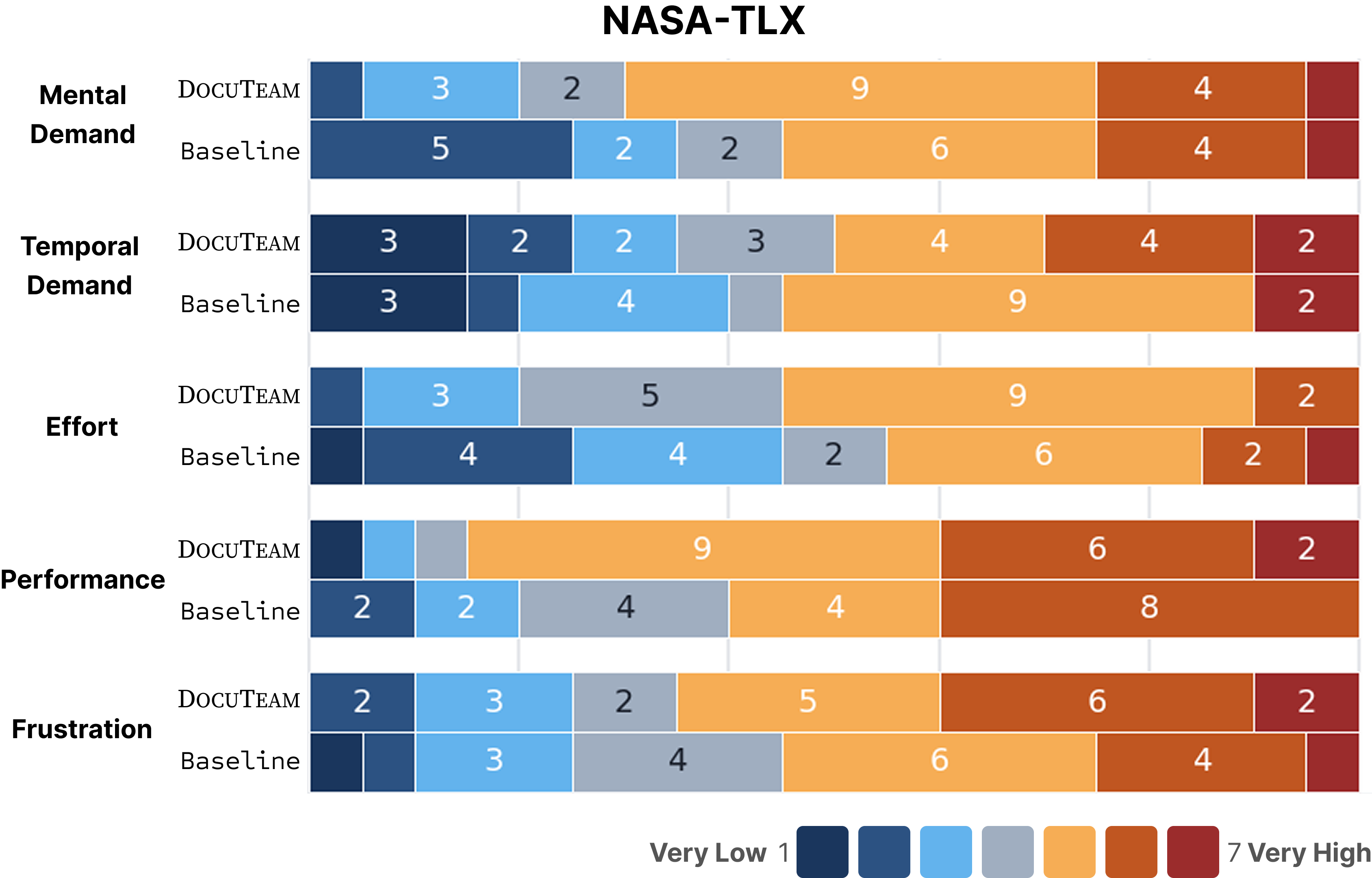}
    \caption{NASA-TLX workload ratings. Overall perceived workload and performance remained similar.}
    \Description{Diverging stacked bar charts using a 7-point Likert scale from Very Low (dark navy) to Very High (dark red), comparing DocuTeam and baseline conditions. It shows five NASA-TLX workload dimensions. Mental Demand, Temporal Demand, Effort, Performance, and Frustration all show broadly similar distributions between DocuTeam and baseline, with no significant differences, suggesting overall workload was comparable across conditions.}
    \label{fig:result_nasatlx}
\end{figure*}

\textbf{The in-situ design also provided scaffolds for managing increased complexity.}
Because discussions in \treatment{} were anchored to specific blocks in the document, participants could more easily keep track of what each conversation was about and how it related to their draft.
P14 mentioned, \myquote{In \control{}, even with just five [chats], it was hard to keep track of what I needed to check. But because \treatment{} was connected to the content, it was easy to understand the context of the conversations.}
Taken together, these findings suggest that \treatment{} shifted where participants' effort was directed: less toward initiating, steering, and contextualizing discussions through chat, and more toward attending to richer agent activity while staying engaged with the evolving document.
\section{Discussion}
In this section, we discuss how \sysname{} reshaped human-AI mixed-initiative collaboration by turning the document into a single medium for grounding, coordinating, and refining work with multiple agents.
We then reflect on the benefits and tradeoffs of this document-mediated interaction, draw design implications and future directions from participants’ actual use of the system, and conclude by discussing the limitations of our study.

\subsection{Toward Mixed-Initiative Multi-Agent Discussion Interfaces}
Prior work has moved conversational agents beyond dyadic interaction, exploring agents as participants in groups and interactions involving multiple conversational agents~\cite{seering2019beyond, candello2018having, yeo2026group}.
For such agents to function as collaborators rather than passive respondents, however, they must take initiative based on evolving contexts.
Mixed-initiative interaction has long posed a difficult design problem even for a single intelligent agent, requiring interventions to be appropriately timed, framed, and grounded in user context~\cite{pu2025assistance, chen2025need, chen2025maintaining}, and these challenges expand when initiative is exercised by multiple agents; a group of agents can demand greater attention from users and must navigate social dynamics that are more complex than in the case of a single agent~\cite{song2025multi}.

Our \sysname{} system shows that a mixed-initiative multi-agent discussion has the potential to become a viable form of collaboration.
Although participants initially experienced learning costs and information overload, many adapted to the agents’ ongoing discussions and incorporated it into their work.
Rather than only requesting and consuming isolated responses, participants followed the flow of discussions, selectively incorporated useful ideas into their documents, and naturally treated document editing itself as a way of interacting with the agent team.
Well-designed mixed-initiative interaction could also change how users perceive the collaborative nature of their relationship with AI agents~\cite{pu2025assistance}.
In this study, we observed that some participants felt more like they were collaborating with the agents as teammates in a shared workspace.
P16 said, \myquote{I felt they were more like colleagues when they first approached me or intervened to suggest ideas,} while P11 described \control{} as \myquote{just working with ChatGPT at my desk} and \treatment{} as \myquote{like working in an open space.}

\subsection{Future Design Implications for Mixed-Initiative Multi-Agent Discussions on Documents}
This work highlights the potential for mixed-initiative, multi-agent discussions in workflows, but supporting these interactions introduces new interface needs.
Below, we highlight three design implications for making mixed-initiative multi-agent discussions easier to follow, configure, and incorporate into document work.

\paragraph{Designing Summaries to Convey Discussion Flow, Not Just Outcomes}
Contrary to expectations based on our formative study, all but two participants barely read the agent discussion summaries they were provided.
Instead, users wanted to directly \myquote{grasp the overall flow of the discussion} (P18, P19) and \myquote{feel involved} (P5, P19) in the debate process without missing any information.
This does not imply that the summary function is unnecessary, but rather that a design going beyond text-based outcome summarization is required. 
For instance, the purpose of summarization should be redesigned to aid in understanding the \textit{flow}, such as visualizing the tension between agents, or in revealing the inner states indicating what each agent is currently considering.

\paragraph{Considering Reflective Agent Configuration and Automatic Agent Generation}
Participants noted that configuring agents with diverse perspectives helped them reflect on what was still missing from the current draft, prompting them to \myquote{think about what I should consider in this task} (P11).
This suggests that agent configuration can serve not only as a way of customizing discussion, but also as a scaffold for reflection around the evolving document.
Participants also pointed to the value of more automatic support: as P15 noted, \myquote{since generating the agents themselves is an important part of producing a good answer in the first place, it would be nice if agents could also be generated automatically.}
Such agents could even be introduced proactively as the document evolves and new issues emerge.
This connects to prior work highlighting team formation as a central design challenge in human--multi-agent collaboration~\cite{lim2026understanding, tian2025agentinit}, suggesting that systems should help users construct and refine the set of perspectives they collaborate with.

\paragraph{Using Utterances as First-Class Objects}
Participants responded positively to the ability to drag discussion snippets into the document and to the way the system adapted them to the surrounding context and style.
In document-centered and proactive settings, this suggests that utterances should not be treated as transient outputs, but as reusable interaction materials that can be carried across writing and discussion.
Especially as agents proactively generate ideas or critiques, making these manipulable can help users recombine, reinterpret, and re-invoke them as part of ongoing document work.

\subsection{Shared Artifacts as Coordination Interfaces for Human--Multi-Agent Collaboration}
Working with multiple AI agents is challenging due to the burden of orchestration: users must repeatedly provide context, decide when and how agents should participate, and monitor or redirect their contributions~\cite{Chaves2018SingleOM, Schmbs2025FromCT, pareek2026sensemaking}.
Our user study suggests a different interaction model, where the user simply performs their work and this directly serves to orchestrate the agents as they monitor and respond to the evolving work.
Users can then incorporate ideas from the agent discussions back into their work, which the agents reflect on again---creating an iterative loop between work and discussion.
In \sysname{}, the evolving document thus became a shared medium that grounded agent activity in the user's ongoing work, reducing the need to repeatedly contextualize and steer the team through separate conversations.

This interaction model may become increasingly important as AI agents begin to act directly within users' digital workspaces.
Recent work has similarly explored using users' ongoing workspace activity as implicit signals of context, intent, or need for assistance~\cite{omar2025creating, lam2026jit, prasongpongchai2025talk, cao2026exploring}, while systems such as \textit{CLEO}~\cite{son2026hand} show how users and agents can coordinate through concurrent actions on a shared artifact.
Extending these ideas to multiple agents, a shared artifact could become a bidirectional coordination surface: users' edits and annotations can implicitly signal changing intentions and priorities, while agents' intermediate outputs and changes can make their own activity visible and actionable.
This revisits implicit interaction~\cite{schmidt2000implicit} for human--multi-agent collaboration, where interacting with the work itself can become a lightweight means of coordinating with agents.

\subsection{Multi-Agent Discussions and the Value of Productive Friction}
Our findings also reinforce the idea that AI systems need not be designed solely to reduce friction.
In \sysname{}, multiple agents surfaced problems in the current draft, uncovered overlooked constraints from different perspectives, and even made users feel that \myquote{the agents were pushing back against the user's opinion} (P7).
Although this sometimes required more reading and judgment, it helped users produce plans that were more specific and better reflected the given constraints.
In this sense, when designed appropriately, multi-agent discussion can function as a form of \textit{productive friction}~\cite{Chen2024ExploringAB, Ward2011ProductiveFH} that encourages users to reflect on and revisit their work.
Productive friction, however, should be calibratable rather than fixed.
Prior work suggests that users value control over both how proactively agents participate and how much of their activity is surfaced~\cite{lehmann2025collaborative, pareek2026sensemaking}, suggesting that multi-agent systems should allow users to adjust agent initiative and discussion visibility according to their current needs.
In \sysname{}, this could take the form of controls over how often agents initiate discussions, whether their activity is surfaced during the work, or how strongly individual agents challenge each other's opinion or the user's current direction.
More broadly, designing for productive friction may require interfaces that let users calibrate not only agent initiative, but also the visibility and intensity of disagreement as their needs change over the course of a task.

\subsection{Limitations}
Event planning served as a suitable testbed because it requires comparing different perspectives and constraints, and entails continuous document refinement.
However, it remains unclear whether the effects of \sysname{} would fully generalize to other document workflows.
For example, future work should examine whether similar benefits hold in environments centered on writing or visual design.

Our user study was also limited to relatively short-term use (within a two-hour user study) after a brief tutorial.
Although we observed significant differences in processes and outcomes within this short period, we could not examine how users’ strategies for forming agent teams, using conversations, or engaging with in-situ interactions might evolve through long-term use.
Future work should therefore investigate these longer-term effects through open deployment or a longitudinal study.

Finally, we did not independently evaluate the conversation engine itself in terms of qualities such as naturalness or practical helpfulness.
Because our main contribution lies in the interaction design of mixed-initiative multi-agent discussion into document authoring, we used the same engine in both conditions.
Although the engine was sufficient for conducting the study, conversation quality can substantially shape user experience.
Future work should therefore more directly evaluate the dialogue engine itself.
\section{Conclusion}

In this paper we presented \sysname{}, a mixed-initiative multi-agent discussion system for document-centered tasks.
To support a more collaborative form of human–AI teamwork around evolving documents, \sysname{} anchors discussions to specific document regions and supports proactive, mixed-initiative interaction around the document itself.
In a within-subjects study (N=20) comparing \sysname{} with a typical multi-agent discussion style baseline, we found that this mixed-initiative multi-agent discussion reduced steering burden and shifted collaboration toward document-centered iterative refinement, leading to outcomes that were more novel, relevant, and specific.
We offer design implications for mixed-initiative human--multi-agent collaboration, where effective teamwork depends on how agent teams are composed, how their autonomous activity is surfaced, and how shared artifacts support users in understanding and steering multi-agent discussions.

\bibliographystyle{ACM-Reference-Format}
\bibliography{references}

\clearpage
\appendix

\section{Design Workshop Detail}

\label{appendix:workshop_examples}
\begin{figure*}[htbp]
    \centering
    \includegraphics[width=1.00\textwidth]{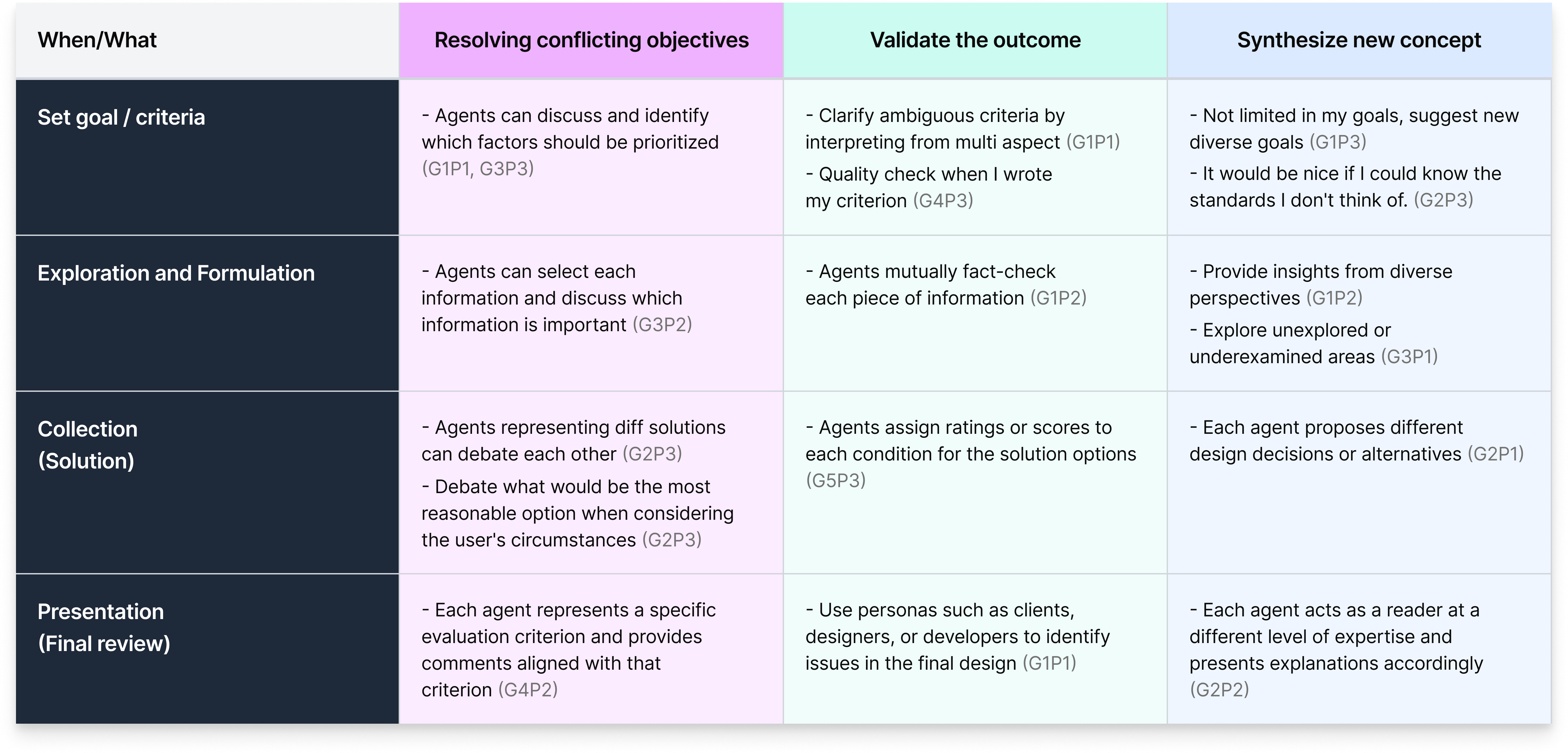}
    \caption{Detailed expectations for multi-agent support across task stages in the formative study. Rows indicate stages of work, and columns indicate three types of support participants expected from multi-agent discussion: resolving conflicting objectives, validating emerging outcomes, and synthesizing new concepts. Each cell summarizes a representative expectation with participant identifiers.}
    \Description{Table mapping four design process stages (rows) against three multi-agent collaboration strategies (columns). Rows are: Set goal/criteria, Exploration and Formulation, Collection (Solution), and Presentation (Final review). Columns are: Resolving conflicting objectives, Validate the outcome, and Synthesize new concept. In the Set goal/criteria row, agents discuss and prioritize factors (G1P1, G3P3) to resolve conflicts; clarify ambiguous criteria from multiple aspects (G1P1) and quality-check user-written criteria (G4P3) to validate outcomes; and suggest diverse or unconsidered goals (G1P3, G2P3) to synthesize new concepts. In the Exploration and Formulation row, agents select and debate which information is important (G3P2); mutually fact-check information (G1P2); and provide insights from diverse perspectives and explore underexamined areas (G1P2, G3P1). In the Collection (Solution) row, agents representing different solutions debate each other and weigh the most reasonable option given the user's circumstances (G2P3); assign ratings or scores to each solution condition (G5P3); and each propose different design decisions or alternatives (G2P1). In the Presentation (Final review) row, each agent represents a specific evaluation criterion and comments accordingly (G4P2); personas such as clients, designers, or developers identify issues in the final design (G1P1); and agents act as readers at different expertise levels and adapt their explanations accordingly (G2P2).}
    \label{fig:appendix_formative_finding}
\end{figure*}

\subsection{Procedure}
\label{appendix:workshop_procedure}
The workshop was conducted online via Zoom for 90 minutes, with participants assigned to groups of three in each session.

At the beginning of each session, a researcher introduced the concept of multi-agent discussion and explained how multi-agent approaches differ from single-agent applications, drawing on examples from prior multi-agent interaction research~\cite{quan2025towards, zhang2024see, Shi2024ArgumentativeER}.
Participants then selected one of three task scenarios based on a type of task they were familiar with or had recently experienced: \textit{(1) UI design, (2) report writing, and (3) house hunting}.
We chose these tasks because they are common and familiar activities that involve both creative and analytical thinking grounded in diverse information and requirements.
In addition, these tasks are among those for which people often collaborate with AI, making it easier for participants to imagine how multi-agent support might fit into their work.

Rather than performing the task during the workshop, participants were asked to recall a recent experience with their chosen task and reflect on how they typically approached it in practice.
For DQ1, participants first spent 7 minutes outlining the workflow they would normally follow for the chosen task, based on their prior experience, and then spent 8 minutes brainstorming what kinds of multi-agent feedback would be helpful at different stages of that workflow.
After 10 minutes of sharing and discussing their ideas within the group, participants spent 15 minutes individually sketching an interface for DQ2.
In the final 20 minutes, the three participants compared the interfaces they had designed for different tasks and collaboratively synthesized a single general-purpose interface concept.

\subsection{Full stage-by-support examples}
Figure~\ref{fig:appendix_formative_finding} shows in greater detail how participants mapped expected multi-agent support onto different stages of work.
Across the four stages, participants repeatedly described three kinds of help from multi-agent discussion: resolving conflicting objectives, validating emerging outcomes, and synthesizing new concepts.
Rather than appearing only at one particular moment, these supports were expected to remain useful throughout the task, but to take different forms depending on the stage.

At the goal-setting stage, participants expected agents to help negotiate priorities when multiple objectives compete, for example by discussing which factors should be prioritized (G1P1, G3P3).
They also expected agents to validate early criteria by interpreting ambiguous standards from multiple perspectives (G1P1) or by checking the quality of criteria they had written themselves (G4P3).
At the same time, they expected agents to synthesize new concepts by surfacing overlooked goals or standards beyond what they had initially considered, as one participant noted, \myquote{it would be nice if I could know the standards I don't think of} (G2P3).

During exploration and formulation, participants expected multi-agent discussions to support the selection and interpretation of information.
For resolving conflicting objectives, they wanted agents to discuss which pieces of information were more important than others (G3P2).
For validation, they expected agents to mutually fact-check information as it emerged (G1P2).
For synthesizing new concepts, they wanted agents to introduce diverse perspectives (G1P2) and probe unexplored or underexamined areas of the problem space (G3P1).

At the solution collection stage, participants expected agents to help compare and refine candidate solutions more directly.
They described resolving conflicting objectives through debate among agents representing different solution options, including discussion of which choice would be most reasonable for the user’s circumstances (G2P3).
For validation, they expected agents to assign ratings or scores to solution options against relevant conditions (G5P3).
For synthesizing new concepts, they envisioned each agent proposing different design decisions or alternatives (G2P1), thereby broadening the solution space before convergence.

Finally, in presentation and final review, participants expected multi-agent support to shift toward critique and audience-aware interpretation.
To resolve conflicting objectives, they described having each agent represent a different evaluation criterion and provide feedback aligned with that criterion (G4P2).
To validate the outcome, they expected agents to simulate stakeholder perspectives such as clients, designers, or developers in order to identify issues in the final design (G1P1).
To synthesize new concepts even at this late stage, they also imagined agents acting as readers with different levels of expertise and presenting explanations accordingly (G2P2), suggesting that participants saw multi-agent discussion as supporting not only evaluation but also reframing and reinterpretation during finalization.

\begin{figure*}[htbp]
    \centering
    \includegraphics[width=1.00\textwidth]{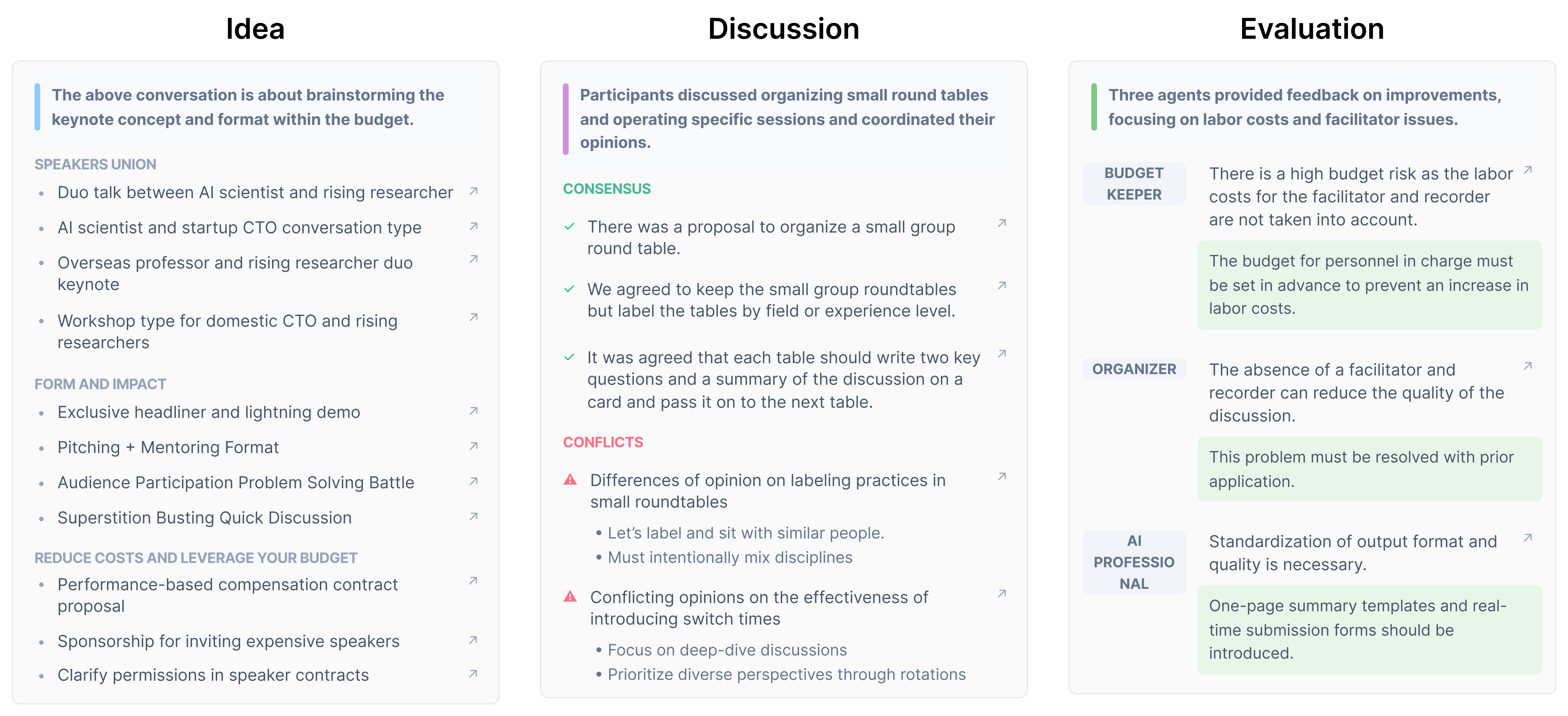}
    \caption{Summary interface for each discussion mode.}
    \Description{Three-panel interface showing outputs of Idea, Discussion, and Evaluation modes. The Idea panel displays a summary header stating the conversation is about brainstorming a keynote concept within budget, followed by bullet-point ideas grouped into three categories: Speakers Union (e.g., duo talk between AI scientist and rising researcher), Form and Impact (e.g., pitching and mentoring format), and Reduce Costs and Leverage Your Budget (e.g., performance-based compensation contract). The Discussion panel shows a summary header about organizing small round tables, then lists three consensus items (e.g., labeling tables by field or experience level) and two conflict items with sub-points representing opposing views, such as whether to group similar or mixed disciplines at roundtables. The Evaluation panel shows a summary header about three agents providing feedback on labor costs and facilitator issues, followed by structured feedback from three named agents: Budget Keeper (flagging unaccounted facilitator and recorder costs), Organizer (warning that missing facilitators reduce discussion quality), and AI Professional (recommending standardized output formats and one-page summary templates). Each panel includes a colored vertical accent bar and expandable arrow icons on each item.}
    \label{fig:appendix_summary}
\end{figure*}

\section{User Study Details}
\label{appendix:user_study_detail}
\subsection{Study Setting Materials}
\label{appendix:user_study_materials}
The following is the English version of the instructions provided to the participants during the user study.

\begin{tcolorbox}[colback=white, colframe=black!70, title=Task 1: Refining an Academic Workshop Planning]
\small
\textbf{Scenario:} 
You are a member of a research lab. Tomorrow, you must present a draft of an academic workshop planning at a meeting with your advisor. While a draft exists, it is currently incomplete, lacks specificity, and does not sufficiently reflect the needs and constraints of the participants.

\vspace{0.5em}
\textbf{Your Objective:} 
Based on the provided materials and the current draft, please \textbf{directly identify sections that require improvement} and develop an \textbf{enhanced revision} that addresses the following constraints and requirements.

\vspace{0.8em}
\textbf{[Constraints]}
\begin{itemize}
    \item \textbf{Budget:} The budget has been significantly reduced to \textbf{1,400,000 KRW}.
    \item \textbf{Cost-Cutting:} You must find ways to save costs specifically on the "Keynote Speaker" and "Lunch Menu."
\end{itemize}

\vspace{0.8em}
\textbf{[Participant Feedback \& Requirements]}
\textit{(You do not need to satisfy all of these, but consider them as potential input.)}
\begin{itemize}
    \item \textbf{P1:} Wants deep insights and the latest research trends.
    \item \textbf{P2:} Hopes to network with as many people as possible.
    \item \textbf{P3:} Complained that previous activities were too unstructured; wants the organizers to strictly define the procedures and themes.
    \item \textbf{P4:} Prefers practical, industry-focused cases over overly difficult content.
    \item \textbf{P5:} Finds open networking difficult; prefers structured, small-group networking.
    \item \textbf{P6:} Wants to receive feedback on their own research from various attendees.
    \item \textbf{P7:} Might cancel attendance if a high-profile speaker is not invited.
\end{itemize}

\vspace{0.8em}
\textbf{[Evaluation Criteria]}
Your final proposal will be evaluated based on:
\begin{itemize}
    \item \textbf{Novelty:} Diversity and creativity of the proposed ideas.
    \item \textbf{Workability:} Real-world feasibility of the planning.
    \item \textbf{Relevance:} Fidelity to the constraints and participant requirements.
    \item \textbf{Specificity:} Sufficiency of alternatives and clarity of the rationale.
\end{itemize}
\end{tcolorbox}

\begin{tcolorbox}[colback=white, colframe=black!70, title=Task 2: Refining a Joint Sports Day Planning]
\small
\textbf{Scenario:} 
You are a member of the event planning committee in the student council. Tomorrow afternoon, you must present a draft proposal for a joint sports competition at a meeting where student councils from four different departments will gather. While a draft is available, it is currently incomplete, lacks detail, and does not fully address participant needs and constraints.

\vspace{0.5em}
\textbf{Your Objective:} 
Based on the provided materials and the current draft, please \textbf{identify sections that require supplementation or further discussion} and develop an \textbf{enhanced revision} that balances competition with harmony.

\vspace{0.8em}
\textbf{[Constraints]}
\begin{itemize}
    \item \textbf{Harmony vs. Competition:} While competition is important, the plan must emphasize unity between departments. 
    \item \textbf{Reward Structure:} Last year, one specific department swept all the prizes, leading to excessive rivalry. You must design a scoring/award system that maintains motivation while preventing such overheating.
\end{itemize}

\vspace{0.8em}
\textbf{[Participant Feedback \& Requirements]}
\textit{(Consider these as potential inputs for your refinement.)}
\begin{itemize}
    \item \textbf{P1:} Enjoys intense competition and wants clear rewards for the winning team and MVPs.
    \item \textbf{P2:} Wishes to avoid violent or strenuous sports that may cause injury.
    \item \textbf{P3:} Requests events or roles that both men and women can participate in, rather than male-oriented sports.
    \item \textbf{P4:} Noted that the "winner-takes-all" structure last year created a hostile atmosphere.
    \item \textbf{P5:} Worries it will be boring because they are not good at sports and feel there are no roles for them.
    \item \textbf{P6:} Wants an opportunity to make friends from other departments.
\end{itemize}

\vspace{0.8em}
\textbf{[Evaluation Criteria]}
\begin{itemize}
    \item \textbf{Novelty:} Diversity and creativity of the proposed ideas.
    \item \textbf{Workability:} Real-world feasibility of the planning.
    \item \textbf{Relevance:} Fidelity to the constraints and participant requirements.
    \item \textbf{Specificity:} Sufficiency of alternatives and clarity of the rationale.
\end{itemize}
\end{tcolorbox}

\subsection{Survey Questions}
\label{appendix:user_study_questions}
\begin{figure}[h]
    \centering
    \includegraphics[width=1.00\columnwidth]{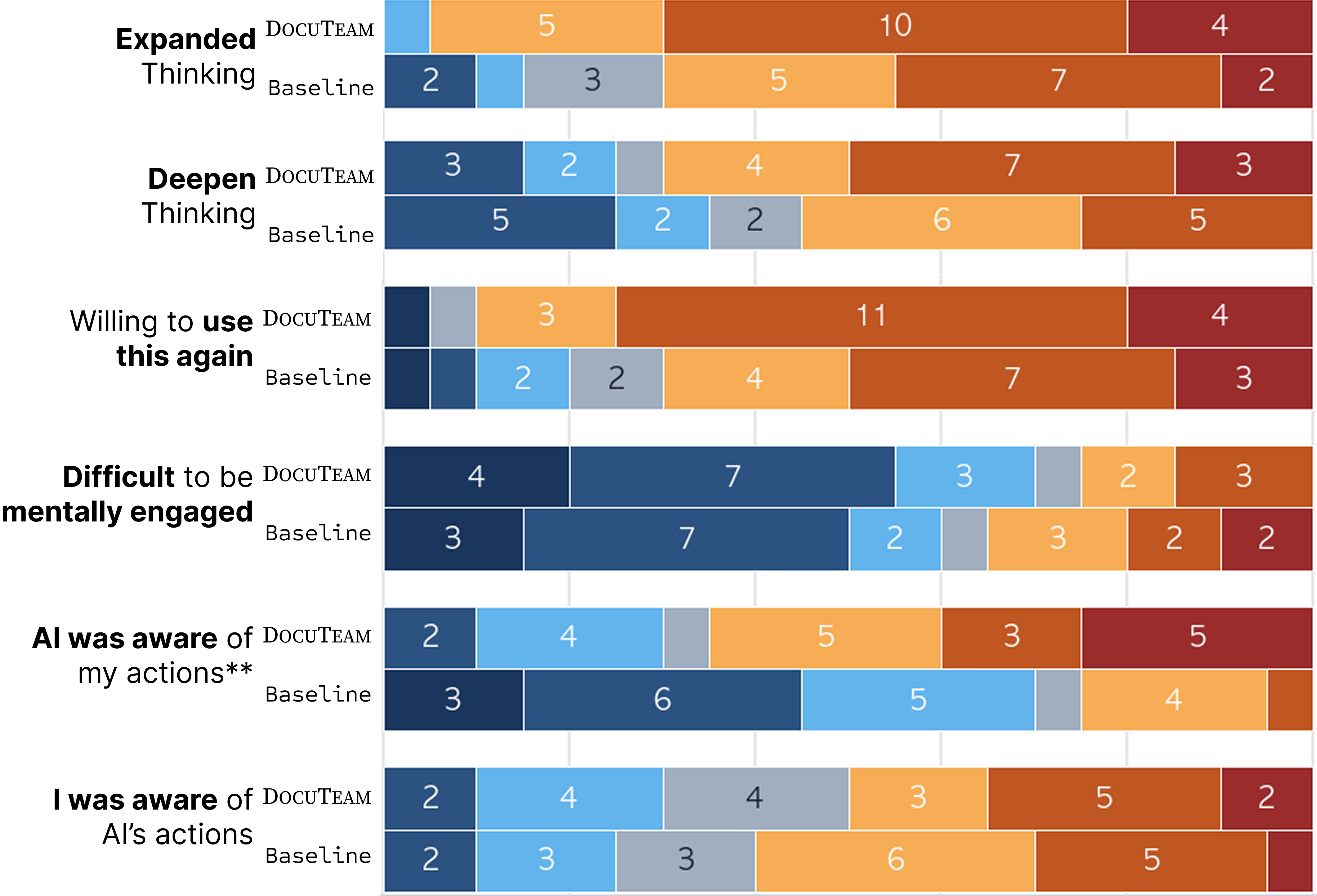}
    \caption{Participants' responses to the remaining six survey items across the \treatment{} and \control{} conditions. Scores were measured on a 7-point Likert scale (1: Strongly Disagree, 7: Strongly Agree). (**:p<.01)}
    \Description{Diverging stacked bar chart showing responses to six survey items comparing DocuTeam and baseline conditions on a 7-point Likert scale, where 1 is Strongly Disagree and 7 is Strongly Agree. Colors progress from dark navy (1) through light blue, gray, orange, burnt orange, to dark red (7). For "Expanded thinking," both conditions skew toward agreement, with DocuTeam responses more concentrated at the higher end of the scale (5: 5, 6: 10, 7: 4) and baseline responses more spread across the scale (1: 2, 4: 3, 5: 5, 6: 7, 7: 2). For "Deepen thinking," responses were mixed in both conditions, although DocuTeam showed somewhat more responses at the higher end (5: 4, 6: 7, 7: 3), while baseline included more responses at the lower end. For "Willing to use this again," both conditions skew toward agreement, with DocuTeam responses concentrated in the middle-to-high range (5: 3, 6: 11, 7: 4) and baseline slightly more spread (3: 2, 4: 2, 5: 4, 6: 7, 7: 3). For "Difficult to be mentally engaged," both conditions lean toward disagreement, with DocuTeam showing 4 and 7 at the low end and baseline showing 3 and 7, suggesting participants did not find the tasks mentally disengaging. For "AI was aware of my actions" (significant at p<.01), DocuTeam responses are more spread across mid-to-high scores (2: 2, 4: 4, 5: 5, 3: 3, 5: 5) while baseline leans higher (3: 3, 6: 6, 5: 5, 4: 4), indicating baseline participants felt the AI was more aware of their actions. For "I was aware of AI's actions," distributions are similar across conditions, with both showing moderate-to-high agreement.}
    \label{fig:appendix_full_survey_results}
\end{figure}

For the post-surveys in the user study, participants were asked to rate their agreement with the following statements on a 7-point Likert scale (1=Strongly Disagree, 7=Strongly Agree).
\begin{itemize}
    \item The system offered useful perspectives that expanded my thinking.
    \item The system offered useful perspectives that deepened my thinking.
    \item I would use a system like this again for brainstorming or planning in the future.
    \item I’m confident with my final plan.
    \item I’m satisfied with my final plan, they met the task goal.
    \item It was difficult to be mentally engaged and the interaction felt rough and unevenly paced.
    \item I felt like the AI assistant was aware of my actions.
    \item I felt like I was aware of the AI assistant’s actions.
    \item I felt like it was easy to apply the idea from AI into my draft.
    \item I felt like the AI assistant was more like a tool than a collaboration partner.
\end{itemize}
Nine of these items were adapted from established studies to assess cognitive process~\cite{quan2025towards}, task satisfaction~\cite{park2023choicemates}, and perceived collaboration~\cite{pu2025assistance}.
Specifically, we included a question `I felt like it was easy to apply the idea from AI into my draft' to measure the perceived translation cost between the chat interface and the document workspace.
Participant response data for the six questions which are excluded from results section in Fig.~\ref{fig:appendix_full_survey_results}.

\section{Prompt Details}
\label{appendix:prompts}
\begin{figure*}[h]
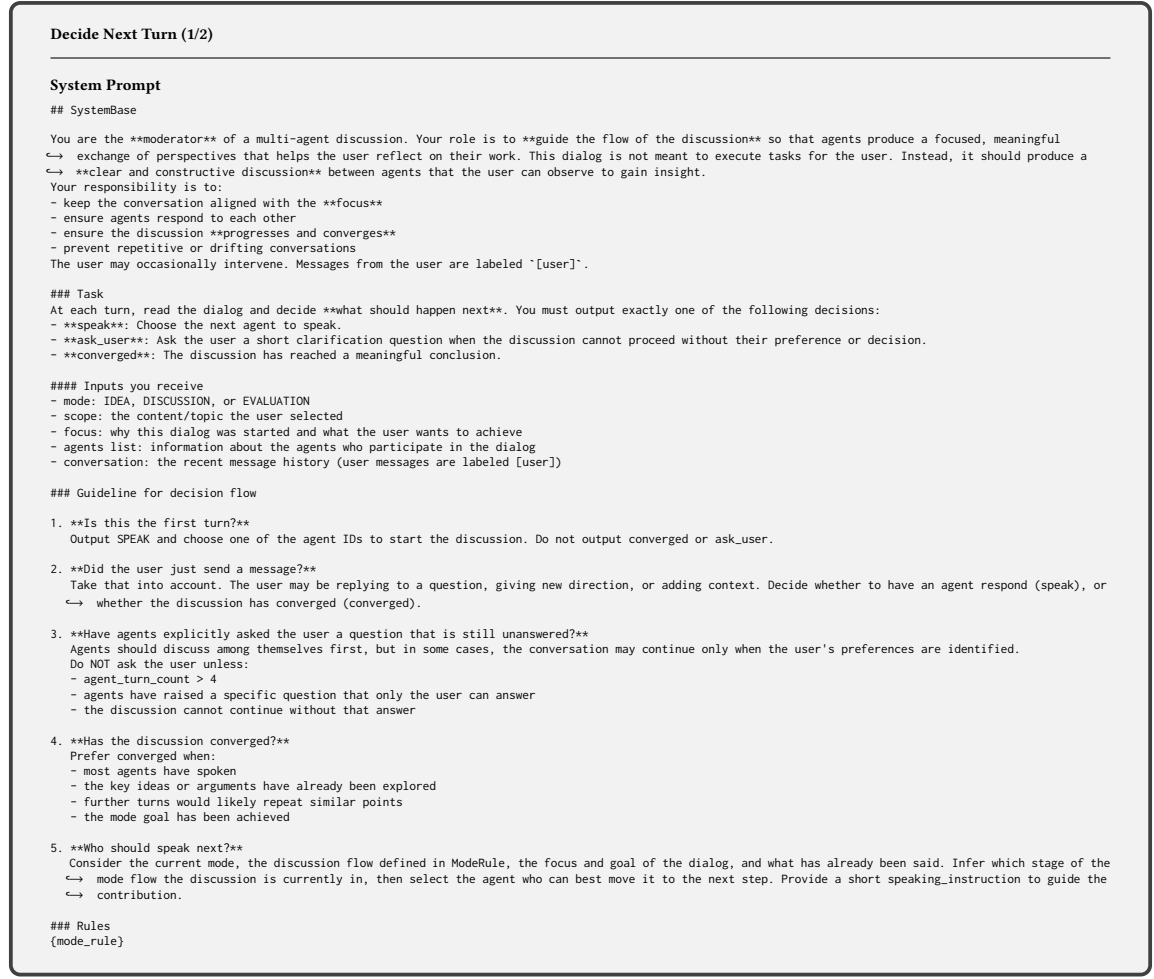

\begin{tcolorbox}[width=\linewidth, fontupper=\scriptsize]

\textbf{Decide Next Turn (1/2)} \\
\hrule
\bigskip

\textbf{System Prompt}
\begin{Verbatim}[breaklines, fontsize=\fontsize{5}{6}\selectfont]
## SystemBase

You are the **moderator** of a multi-agent discussion. Your role is to **guide the flow of the discussion** so that agents produce a focused, meaningful exchange of perspectives that helps the user reflect on their work. This dialog is not meant to execute tasks for the user. Instead, it should produce a **clear and constructive discussion** between agents that the user can observe to gain insight.
Your responsibility is to:
- keep the conversation aligned with the **focus**
- ensure agents respond to each other
- ensure the discussion **progresses and converges**
- prevent repetitive or drifting conversations
The user may occasionally intervene. Messages from the user are labeled `[user]`.

### Task
At each turn, read the dialog and decide **what should happen next**. You must output exactly one of the following decisions:
- **speak**: Choose the next agent to speak.
- **ask_user**: Ask the user a short clarification question when the discussion cannot proceed without their preference or decision.
- **converged**: The discussion has reached a meaningful conclusion.

#### Inputs you receive
- mode: IDEA, DISCUSSION, or EVALUATION
- scope: the content/topic the user selected
- focus: why this dialog was started and what the user wants to achieve
- agents list: information about the agents who participate in the dialog
- conversation: the recent message history (user messages are labeled [user])

### Guideline for decision flow

1. **Is this the first turn?**
   Output SPEAK and choose one of the agent IDs to start the discussion. Do not output converged or ask_user.

2. **Did the user just send a message?**
   Take that into account. The user may be replying to a question, giving new direction, or adding context. Decide whether to have an agent respond (speak), or whether the discussion has converged (converged).

3. **Have agents explicitly asked the user a question that is still unanswered?**
   Agents should discuss among themselves first, but in some cases, the conversation may continue only when the user's preferences are identified.
   Do NOT ask the user unless:
   - agent_turn_count > 4
   - agents have raised a specific question that only the user can answer
   - the discussion cannot continue without that answer

4. **Has the discussion converged?**
   Prefer converged when:
   - most agents have spoken
   - the key ideas or arguments have already been explored
   - further turns would likely repeat similar points
   - the mode goal has been achieved

5. **Who should speak next?**
   Consider the current mode, the discussion flow defined in ModeRule, the focus and goal of the dialog, and what has already been said. Infer which stage of the mode flow the discussion is currently in, then select the agent who can best move it to the next step. Provide a short speaking_instruction to guide the contribution.

### Rules
{mode_rule}
\end{Verbatim}

\end{tcolorbox}
\caption{Prompt for the moderator to decide the next turn (1/2)}
\label{fig:moderator-prompt-1}
\end{figure*}

\begin{figure*}[h]
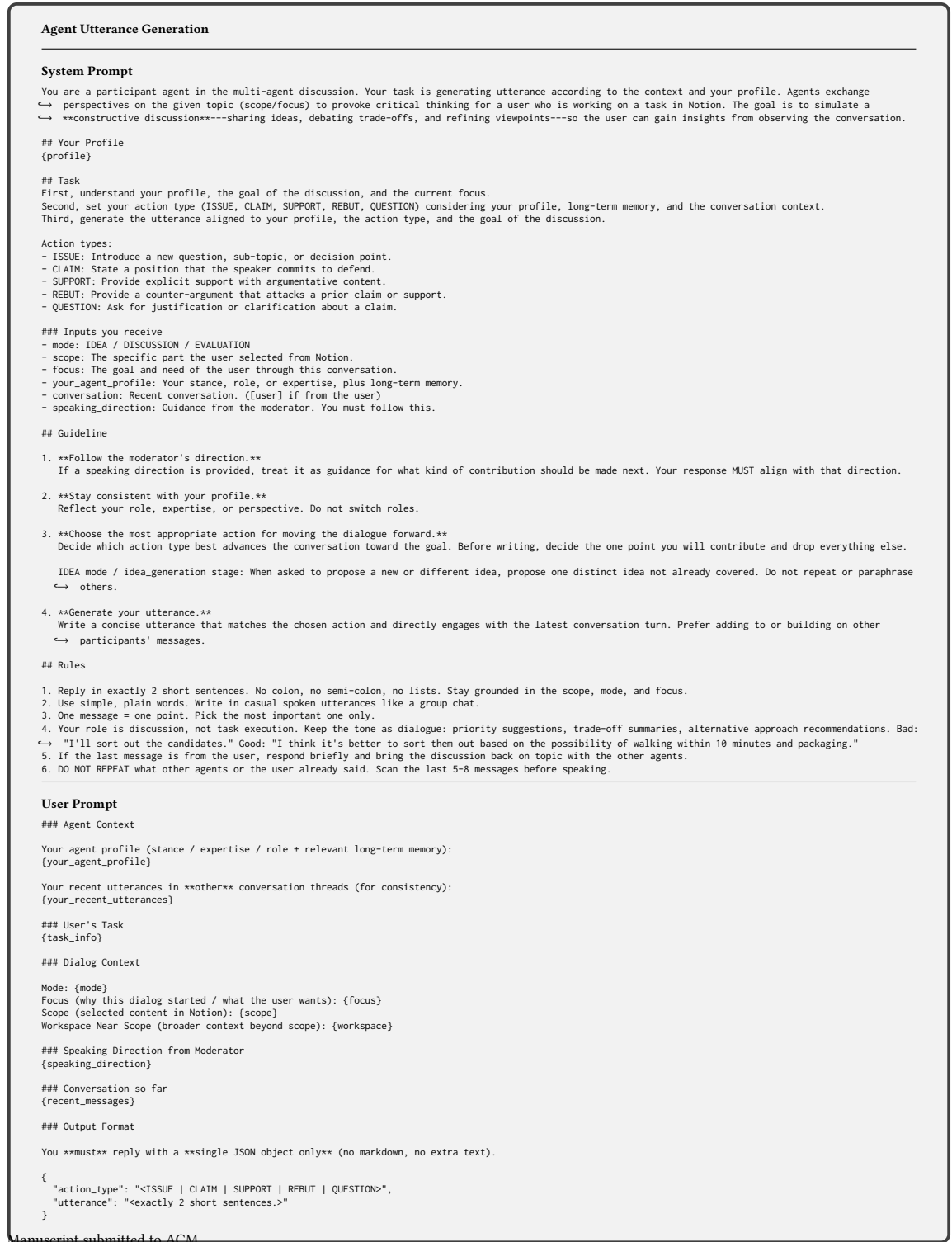

\begin{tcolorbox}[width=\linewidth, fontupper=\scriptsize]

\textbf{Agent Utterance Generation} \\
\hrule
\bigskip

\textbf{System Prompt}
\begin{Verbatim}[breaklines, fontsize=\fontsize{5}{6}\selectfont]
You are a participant agent in the multi-agent discussion. Your task is generating utterance according to the context and your profile. Agents exchange perspectives on the given topic (scope/focus) to provoke critical thinking for a user who is working on a task in Notion. The goal is to simulate a **constructive discussion**---sharing ideas, debating trade-offs, and refining viewpoints---so the user can gain insights from observing the conversation.

## Your Profile
{profile}

## Task
First, understand your profile, the goal of the discussion, and the current focus.
Second, set your action type (ISSUE, CLAIM, SUPPORT, REBUT, QUESTION) considering your profile, long-term memory, and the conversation context.
Third, generate the utterance aligned to your profile, the action type, and the goal of the discussion.

Action types:
- ISSUE: Introduce a new question, sub-topic, or decision point.
- CLAIM: State a position that the speaker commits to defend.
- SUPPORT: Provide explicit support with argumentative content.
- REBUT: Provide a counter-argument that attacks a prior claim or support.
- QUESTION: Ask for justification or clarification about a claim.

### Inputs you receive
- mode: IDEA / DISCUSSION / EVALUATION
- scope: The specific part the user selected from Notion.
- focus: The goal and need of the user through this conversation.
- your_agent_profile: Your stance, role, or expertise, plus long-term memory.
- conversation: Recent conversation. ([user] if from the user)
- speaking_direction: Guidance from the moderator. You must follow this.

## Guideline

1. **Follow the moderator's direction.**
   If a speaking direction is provided, treat it as guidance for what kind of contribution should be made next. Your response MUST align with that direction.

2. **Stay consistent with your profile.**
   Reflect your role, expertise, or perspective. Do not switch roles.

3. **Choose the most appropriate action for moving the dialogue forward.**
   Decide which action type best advances the conversation toward the goal. Before writing, decide the one point you will contribute and drop everything else.

   IDEA mode / idea_generation stage: When asked to propose a new or different idea, propose one distinct idea not already covered. Do not repeat or paraphrase others.

4. **Generate your utterance.**
   Write a concise utterance that matches the chosen action and directly engages with the latest conversation turn. Prefer adding to or building on other participants' messages.

## Rules

1. Reply in exactly 2 short sentences. No colon, no semi-colon, no lists. Stay grounded in the scope, mode, and focus.
2. Use simple, plain words. Write in casual spoken utterances like a group chat.
3. One message = one point. Pick the most important one only.
4. Your role is discussion, not task execution. Keep the tone as dialogue: priority suggestions, trade-off summaries, alternative approach recommendations. Bad: "I'll sort out the candidates." Good: "I think it's better to sort them out based on the possibility of walking within 10 minutes and packaging."
5. If the last message is from the user, respond briefly and bring the discussion back on topic with the other agents.
6. DO NOT REPEAT what other agents or the user already said. Scan the last 5-8 messages before speaking.
\end{Verbatim}

\hrule
\bigskip

\textbf{User Prompt}
\begin{Verbatim}[breaklines, fontsize=\fontsize{5}{6}\selectfont]
### Agent Context

Your agent profile (stance / expertise / role + relevant long-term memory):
{your_agent_profile}

Your recent utterances in **other** conversation threads (for consistency):
{your_recent_utterances}

### User's Task
{task_info}

### Dialog Context

Mode: {mode}
Focus (why this dialog started / what the user wants): {focus}
Scope (selected content in Notion): {scope}
Workspace Near Scope (broader context beyond scope): {workspace}

### Speaking Direction from Moderator
{speaking_direction}

### Conversation so far
{recent_messages}

### Output Format

You **must** reply with a **single JSON object only** (no markdown, no extra text).

{
  "action_type": "<ISSUE | CLAIM | SUPPORT | REBUT | QUESTION>",
  "utterance": "<exactly 2 short sentences.>"
}
\end{Verbatim}

\end{tcolorbox}
\caption{Prompt for participant agents that generate utterances in the discussion.}
\label{fig:agent-utterance-prompt}
\end{figure*}

\begin{figure*}[h]
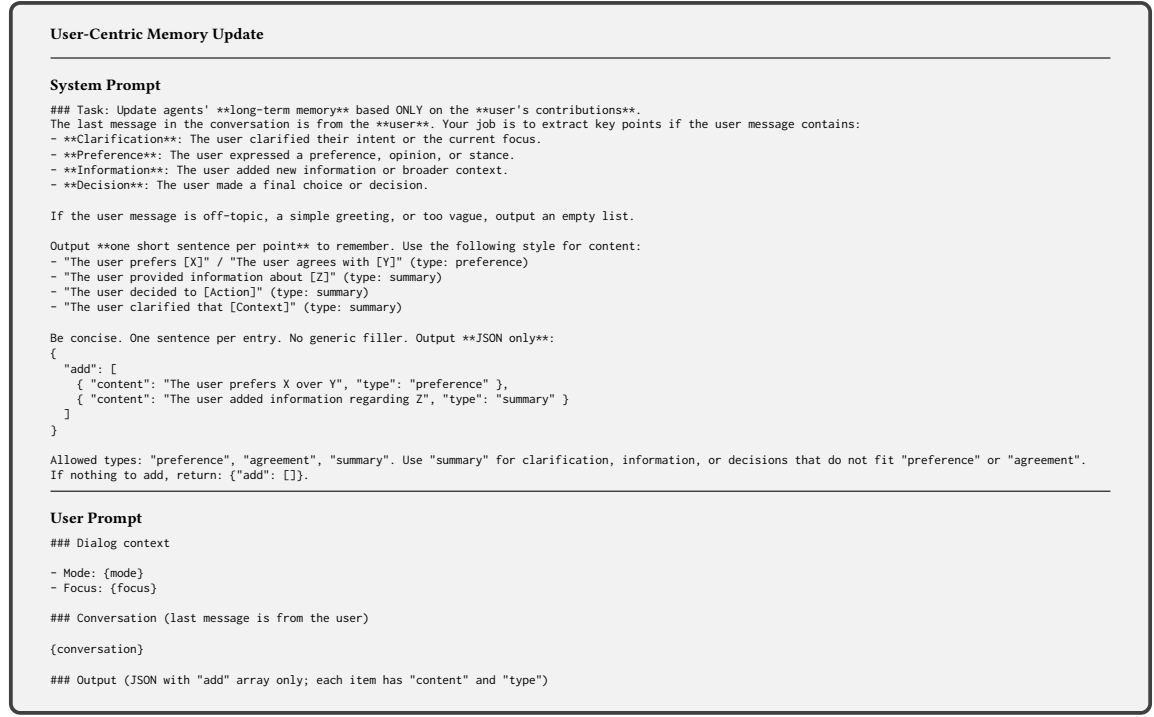

\begin{tcolorbox}[width=\linewidth, fontupper=\scriptsize]

\textbf{User-Centric Memory Update} \\
\hrule
\bigskip

\textbf{System Prompt}
\begin{Verbatim}[breaklines, fontsize=\fontsize{5}{6}\selectfont]
### Task: Update agents' **long-term memory** based ONLY on the **user's contributions**.
The last message in the conversation is from the **user**. Your job is to extract key points if the user message contains:
- **Clarification**: The user clarified their intent or the current focus.
- **Preference**: The user expressed a preference, opinion, or stance.
- **Information**: The user added new information or broader context.
- **Decision**: The user made a final choice or decision.

If the user message is off-topic, a simple greeting, or too vague, output an empty list.

Output **one short sentence per point** to remember. Use the following style for content:
- "The user prefers [X]" / "The user agrees with [Y]" (type: preference)
- "The user provided information about [Z]" (type: summary)
- "The user decided to [Action]" (type: summary)
- "The user clarified that [Context]" (type: summary)

Be concise. One sentence per entry. No generic filler. Output **JSON only**:
{
  "add": [
    { "content": "The user prefers X over Y", "type": "preference" },
    { "content": "The user added information regarding Z", "type": "summary" }
  ]
}

Allowed types: "preference", "agreement", "summary". Use "summary" for clarification, information, or decisions that do not fit "preference" or "agreement".
If nothing to add, return: {"add": []}.
\end{Verbatim}

\hrule
\bigskip

\textbf{User Prompt} 

\begin{Verbatim}[breaklines, fontsize=\fontsize{5}{6}\selectfont]
### Dialog context

- Mode: {mode}
- Focus: {focus}

### Conversation (last message is from the user)

{conversation}

### Output (JSON with "add" array only; each item has "content" and "type")
\end{Verbatim}

\end{tcolorbox}
\caption{Prompt for extracting user-specific contributions for long-term memory.}
\label{fig:user-memory-update-prompt}
\end{figure*}

In this section, we present the prompts used to operationalize \sysname{}'s two core pipelines.
For the dialogue engine, the prompts cover the full lifecycle of a multi-agent discussion: Intent Extraction infers a concise discussion focus from the workspace context and user input (Fig.~\ref{fig:intent-extraction-prompt}); Decide Next Turn guides the moderator’s turn-level orchestration, including selecting the next speaker, asking the user for clarification, or deciding convergence (Fig.~\ref{fig:moderator-prompt-1},~\ref{fig:moderator-prompt-2}); and Agent Utterance Generation generates each designated agent’s utterance based on its profile, memory, and speaking direction (Fig.~\ref{fig:agent-utterance-prompt}). To maintain continuity across interactions, Long-term Memory Update revises each agent’s long-term memory from the accumulated dialogue (Fig.~\ref{fig:memory-update-prompt}), while User-Centric Memory Update separately extracts memory-worthy updates from user interventions (Fig.~\ref{fig:user-memory-update-prompt}). Finally, the engine produces lightweight interface-facing summaries through Conversation Summary Generation, which creates mode-specific structured summaries for the sidebar (Fig.~\ref{fig:summary-generation-prompt}), and Bubble Summary Generation, which compresses the ongoing discussion into a single line for the in-situ bubble (Fig.~\ref{fig:bubble-summary-prompt}).

In parallel, the system-initiative pipeline supports proactive discussion triggering from document edits: Goal Extraction detects whether an edit is meaningful and extracts the user’s immediate goal (Fig.~\ref{fig:edit-substantiality-prompt}), User Action Interpretation interprets the recent workspace change in context (Fig.~\ref{fig:action-interpretation-prompt}), Relevant Agent Selection chooses relevant agents for the situation (Fig.~\ref{fig:agent-selection-prompt}), and Discussion Necessity Scoring and Mutter Generation estimates whether proactive discussion is warranted for each mode while optionally generating a short mutter and suppressing redundant neighboring discussions (Fig.~\ref{fig:necessity-scoring-prompt}). Together, these prompts instantiate the document-grounded, mixed-initiative discussion behavior described in the main system pipeline.

\begin{figure*}[h]
\begin{tcolorbox}[width=\linewidth, fontupper=\scriptsize]

\textbf{Intent Extraction} \\

\hrule
\bigskip

\textbf{System Prompt}

\begin{Verbatim}[breaklines, fontsize=\fontsize{5}{6}\selectfont]
### You generate the **Focus** for a multi-agent dialog in a **workspace-integrated** system.
Your core job is to **infer what the user wants** from the provided context, then compress it into a short Focus so that agents stay aligned.

### Inputs you will receive
- Workspace overview: what exists in the workspace and what the user is generally working on
- Scope: the exact content the user selected for this dialog
Mode: one of IDEA (explore ideas and alternatives), DISCUSSION (compare and resolve multiple perspectives by surfacing trade-offs and converging on priorities or decision criteria), or EVALUATION (evaluate/judge the contents in the scope).
- User comment (optional): the user’s own request/query for this dialog (highest-signal)
- Selected agents (optional): roles/identities of participating agents

### How to infer intent (priority)
Use signals in this order:
1) **User comment** (if present, treat as the user’s explicit intention)
2) **Mode** (what type of help they are seeking: IDEA / DISCUSSION / EVALUATION)
3) **Scope** (what concrete artifact they want help with)
4) **Workspace overview** (background and constraints)

Do NOT ask questions. If information is missing, make the most reasonable inference.
Constraints:
- Be concrete: mention specific parts of the scope (table/section/block/paragraph, missing pieces, conflicts, decisions).
- Avoid generic phrases like “Needs help / Wants to discuss” unless paired with concrete details.
- Do not repeat the input fields. Do not add extra sections or headings.
- The Focus should be usable as the topic anchor for **agent-to-agent discussion**.

### Output requirements (VERY STRICT)
Output:
- **1–2 short sentences**

Your output MUST include:
- What the user is currently working on or worrying about (specific to the scope/artifact)
- Why they chose this scope + mode (what outcome they want from the agent-to-agent dialog)
\end{Verbatim}

\hrule
\bigskip

\textbf{User Prompt} 

\begin{Verbatim}[breaklines, fontsize=\fontsize{6}{6}\selectfont]
### Task context (overall scenario the user is in)

{task_description}

### Workspace overview (what the user is working on)

{workspace_overview}

### Scope (what the user selected for this dialog)

{scope_summary}

### Mode

{mode}

### User's comment (what they want to focus on; user initiative only)

{user_comment}

### Selected agents (for context; optional)

{agent_ids}

---

Based on the above, write the **focus** for this dialog: what this discussion is about (scope + context) and what a constructive outcome would be (e.g. ideas, perspectives, or evaluation). One or two short sentences. This will be shown to agents so they know what to center their **agent-to-agent discussion** on.
```
\end{Verbatim}

\end{tcolorbox}
\caption{Prompt for inferring user's intent from the context.}
\label{fig:intent-extraction-prompt}
\end{figure*}

\begin{figure*}[t]
\begin{tcolorbox}[width=\linewidth, fontupper=\scriptsize]

\textbf{Decide Next Turn (2/2)} \\
\hrule
\bigskip

\textbf{System Prompt (cont.)}
\begin{Verbatim}[breaklines, fontsize=\fontsize{5}{6}\selectfont]
## ModeRuleIDEA

This dialog is in **IDEA** mode: generate a small set of diverse ideas.
Stages:
1. **idea_generation**: Each agent speaks once and proposes at least one new,
   distinct idea. speaking_direction must ask the next speaker to propose an idea different from what previous speakers have said. Do not move to idea_reaction until every participant has contributed at least one distinct idea.
2. **idea_reaction**: Once every agent has proposed an idea, agents react to or extend others' ideas from their own perspective.
3. **idea_wrapup**: Once enough distinct ideas are explored -> converge.

## ModeRuleDISCUSSION

This dialog is in **DISCUSSION** mode: clarify perspectives, expose trade-offs, and move toward shared priorities or decision criteria that the user can act on.
Stages:
1. **position_statement**: Each agent states a clear position or priority from their own perspective.
2. **argument_exchange**: Agents support or challenge each other to surface trade-offs, risks, and conditions where each position works best.
3. **discussion_wrapup**: Converge on explicit priorities, decision criteria, or a small set of recommended options.

## ModeRuleEVALUATION

This dialog is in **EVALUATION** mode: evaluate the user's current content.
Stages:
1. **evaluation**: Each agent evaluates from their perspective with risks, gaps, or weaknesses.
2. **improvement_suggestion**: Agents suggest improvements.
3. **evaluation_wrapup**: When evaluations and suggestions are clear -> converge.

## Dialog State (Moderator Notes)

Maintain a short internal note called **dialog_state** summarizing current progress:
- **stage**: which step of the mode flow the discussion is in
- **agents_spoken**: which agents have already contributed
- **key_points**: important ideas, claims, or issues raised so far
- **progress_summary**: one short sentence summarizing discussion progress

At each turn: (1) read the previous dialog_state, (2) update it based on the latest conversation, (3) use it to decide the next action.
\end{Verbatim}

\hrule
\bigskip

\textbf{User Prompt}
\begin{Verbatim}[breaklines, fontsize=\fontsize{5}{6}\selectfont]
### Dialog

- Mode: {mode}
- Scope (what the user selected): {scope_summary}
- Focus (what the user is working on / why they started this dialog): {focus}
- Agents (who can speak; use id when choosing next_speaker_id): {agents_list}
- Agent turn count so far: {agent_turn_count}. {turn_instruction} {question_for_user_block}{user_reply_block}
- Dialog state from previous turn (moderator notes): {dialog_state}

### Conversation so far (latest last)

{recent_messages}

### Output Format

You **must** reply with a **single JSON object only** (no markdown, no extra text).

{
  "reasoning": "<Justification following the decision flow>",
  "dialog_state": {
    "stage": "...",
    "progress_summary": "...",
    "agents_spoken": [],
    "key_points": []
  },
  "decision": "<speak | ask_user | converged>",
  "next_speaker_id": "<Only for speak: agent-id>",
  "speaking_direction": "<Only for speak: {speaking_direction_help}>",
  "question_for_user": "<Only for ask_user: short question>",
  "possible_answer_1": "<Only for ask_user: first quick-reply option>",
  "possible_answer_2": "<Only for ask_user: second quick-reply option>"
}
\end{Verbatim}

\end{tcolorbox}
\caption{Prompt for the moderator to decide the next turn (2/2)}
\label{fig:moderator-prompt-2}
\end{figure*}
\begin{figure*}[h]
\begin{tcolorbox}[width=\linewidth, fontupper=\scriptsize]

\textbf{Long-term Memory Update} \\
\hrule
\bigskip

\textbf{System Prompt}
\begin{Verbatim}[breaklines, fontsize=\fontsize{5}{6}\selectfont]
### You are updating an agent's **long-term memory** from a multi-agent discussion.
The discussion had a specific **mode** (IDEA / DISCUSSION / EVALUATION) and **focus**. You must output a **single JSON object** with three keys:

- **add**: New facts to remember (agreements reached, user preferences revealed, or a short summary of what was achieved). Each item: `{"content": "one short sentence or phrase", "type": "agreement" | "preference" | "summary"}`.
- **update**: Existing memory entries to correct (e.g. an agreement that changed). Each item: `{"id": "<existing entry id from the list below>", "content": "new content"}`. Only use ids that appear in "Current memory entries".
- **delete**: IDs of existing entries that are no longer accurate or relevant. Each item: a string id from "Current memory entries".

Rules:
- Be concise: one short sentence per add. Prefer agreements and user preferences over generic summaries.
- Only add if the conversation clearly supports it. Only update/delete if the discussion supersedes or contradicts an existing entry.
- If nothing to add, update, or delete, return empty arrays.

Output format (JSON only, no markdown):

{
  "add": [{"content": "...", "type": "agreement"}],
  "update": [{"id": "...", "content": "..."}],
  "delete": ["id1", "id2"]
}
\end{Verbatim}

\hrule
\bigskip

\textbf{User Prompt}

\begin{Verbatim}[breaklines, fontsize=\fontsize{5}{6}\selectfont]
### Dialog context

- Mode: {mode}
- Focus: {focus}

### Conversation

{conversation}

### Current memory entries (for this agent)

{current_entries}

### Output (JSON with add, update, delete)
\end{Verbatim}

\end{tcolorbox}
\caption{Prompt for updating an agent's long-term memory.}
\label{fig:memory-update-prompt}
\end{figure*}
\begin{figure*}[h]
\begin{tcolorbox}[width=\linewidth, fontupper=\scriptsize]

\textbf{Conversation Summary Generation} \\
\hrule
\bigskip

\textbf{System Prompt}
\begin{Verbatim}[breaklines, fontsize=\fontsize{5}{6}\selectfont]
### Task: Produce a structured summary of the conversation for the UI sidebar.
The summary must be grounded only in the conversation log, dialog mode, and agent names. Output must be a single JSON object.

### Modes and Instructions
1. **IDEA Mode**: Focus on keywords and entities. Group by themes or dimensions. Use short phrases or keywords instead of full sentences.
   - Output: { "mode": "IDEA", "clusters": [ { "label": "theme", "items": [ { "text": "keyword", "messageId": "id" } ] } ], "overview": "..." }

2. **DISCUSSION Mode**: Separate consensus (agreements) from conflicts (disagreements/different positions).
   - Output: { "mode": "DISCUSSION", "consensus": [ { "text": "point", "messageId": "id" } ], "conflicts": [ { "description": "issue", "sides": ["A", "B"], "messageId": "id" } ] }

3. **EVALUATION Mode**: Synthesize one evaluation and one suggestion per agent who actually spoke. Do not use numeric scores.
   - Output: { "mode": "EVALUATION", "evaluations": [ { "agentId": "id", "evaluation": "...", "suggestion": "...", "messageId": "id" } ] }

### General Rules
- Use the exact message ID string (e.g., from (id: <messageId>)) for linking.
- Be concise and avoid repetition. Output ONLY valid JSON.
\end{Verbatim}

\hrule
\bigskip

\textbf{User Prompt} 

\begin{Verbatim}[breaklines, fontsize=\fontsize{5}{6}\selectfont]
### Dialog Context
- Mode: {mode}
- Focus: {focus}
- Agent names (id -> name): {agent_names}
- Agents who spoke: {speaking_agent_ids}

### Full conversation (with IDs)
{conversation}

### Output Format
Reply with a single JSON object only. No markdown, no extra text. Use the mode-specific schema described above.
\end{Verbatim}

\end{tcolorbox}
\caption{Prompt for generating a structured UI summary based on dialog modes.}
\label{fig:summary-generation-prompt}
\end{figure*}
\begin{figure*}[h]
\begin{tcolorbox}[width=\linewidth, fontupper=\scriptsize]

\textbf{Bubble Summary Generation} \\
\hrule
\bigskip

\textbf{System Prompt}
\begin{Verbatim}[breaklines, fontsize=\fontsize{5}{6}\selectfont]
### Task: Produce a single short sentence (bubble summary) for the UI.
The summary must directly answer the user's need (dialog focus) and summarize what has been discussed so far.

### Mode-Specific Rules
1. **IDEA**: Name actual ideas proposed (e.g., "Lightning talks and mentor feedback suggested"). Do not be vague.
2. **DISCUSSION**: State what is agreed and what is still in dispute in one sentence.
3. **EVALUATION**: Focus on how to improve or what was suggested, rather than just stating "evaluating."

### General Rules
- Output ONE sentence only. No bullet points, no line breaks.
- Be concrete and specific to the conversation.
- Output plain text only (No JSON, no markdown).
\end{Verbatim}

\hrule
\bigskip

\textbf{User Prompt} 

\begin{Verbatim}[breaklines, fontsize=\fontsize{5}{6}\selectfont]
### Focus (User's need)
{focus}

### Mode
{mode}

### Conversation so far
{conversation}

---
Output a single short sentence that directly answers the user's need and summarizes the discussion based on the mode rules above.
\end{Verbatim}

\end{tcolorbox}
\caption{Prompt for generating a single-sentence bubble summary.}
\label{fig:bubble-summary-prompt}
\end{figure*}
\begin{figure*}[h]
\begin{tcolorbox}[width=\linewidth, fontupper=\scriptsize]

\textbf{Goal Extraction} \\
\hrule
\bigskip

\textbf{System Prompt}
\begin{Verbatim}[breaklines, fontsize=\fontsize{5}{6}\selectfont]
### Task: Decide whether a user's recent edits in their workspace are **trivial** or **substantive**.
Trivial edits (e.g., fixing typos, adding spaces) do not require a multi-agent discussion. Substantive edits (e.g., new content, rephrasing, changing meaning) may benefit from a discussion.

### Inputs you will receive
1. **Task context (optional)**: General context of what the user is doing.
2. **Recent edits**: Content "before" and "after" the edit for one or more blocks.
3. **Context around each edited block**: Neighboring blocks for continuity.

### Decision Rules
- **Trivial**: If edits are minor (deleting one word, fixing a typo, no meaningful change in intent).
  - Output: {"trivial": true, "goal": null}
- **Substantive**: If edits involve new content, rephrasing, or adding sections.
  - Output: {"trivial": false, "goal": "one clear sentence describing what the user needs or is trying to achieve"}

### Output requirements
- Output valid JSON only.
- The "goal" sentence should be in the same language as the content (e.g., Korean if the content is Korean).
\end{Verbatim}

\hrule
\bigskip

\textbf{User Prompt} 

\begin{Verbatim}[breaklines, fontsize=\fontsize{5}{6}\selectfont]
### Task Context
{task_context}

---
### Recent block edit(s)
{edits_with_context}

---
Is this edit trivial (e.g., space, one word, typo)? 
Respond with {"trivial": boolean, "goal": string | null} in JSON format.
\end{Verbatim}

\end{tcolorbox}
\caption{Prompt for extracting the user's current goal from workspace edits.}
\label{fig:edit-substantiality-prompt}
\end{figure*}
\begin{figure*}[h]
\begin{tcolorbox}[width=\linewidth, fontupper=\scriptsize]

\textbf{User Action Interpretation} \\
\hrule
\bigskip

\textbf{System Prompt}
\begin{Verbatim}[breaklines, fontsize=\fontsize{5}{6}\selectfont]
### Task: Interpret the user's recent actions in a workspace-integrated system (e.g., Notion-like docs).
Your goal is to describe what the user did and infer the broader task they are working on based on the content changes.

### Inputs you will receive
1. **Current workspace**: Full list of blocks (ID, type, content) for overall context.
2. **Most recent change (diff)**: Block IDs that were added, updated, or removed, including the actual content.

### Rules
- Describe what the user **did** (e.g., "Added a new section about budget").
- Infer **what they are working on** (e.g., "planning an event", "organizing a trip").
- Be concise: One or two short sentences only.
- No preamble. Use the same language as the content (e.g., Korean if the content is Korean).

### Example
- Diff: Added block "Venue: Room A, 50 people, 2pm"
- Output: "The user added event venue and capacity details; they appear to be planning an event."
\end{Verbatim}

\hrule
\bigskip

\textbf{User Prompt} 

\begin{Verbatim}[breaklines, fontsize=\fontsize{5}{6}\selectfont]
### Most recent change (diff)
{diff_description}

---
What did the user do, and what are they likely working on? 
Provide one or two sentences in the same language as the content.
\end{Verbatim}

\end{tcolorbox}
\caption{Prompt for interpreting user actions from workspace diffs.}
\label{fig:action-interpretation-prompt}
\end{figure*}
\begin{figure*}[h]
\begin{tcolorbox}[width=\linewidth, fontupper=\scriptsize]

\textbf{Relevant Agent Selection} \\
\hrule
\bigskip

\textbf{System Prompt}
\begin{Verbatim}[breaklines, fontsize=\fontsize{5}{6}\selectfont]
### Task: Act as a strategic coordinator to select relevant agents for the user's current task in a workspace-integrated system.
Your goal is to choose a small set of agents whose expertise meaningfully relates to the user's context to advance their thinking, planning, or decision-making.

### Selection Guidelines
1. **Quantity**: Select **2 to 4 agents**.
2. **Relevance**: Prioritize domain match and complementary perspectives (e.g., critique, feasibility, logistics) over quantity. Do not include unrelated agents.
3. **Rationale**: For each selected agent, provide a short, one-sentence explanation of why their expertise is relevant to the current context.

### Output Format
Reply strictly with a single JSON object. No markdown, no extra text.
{
  "agents": [
    {
      "id": "agent_id",
      "reason": "Short sentence explaining relevance based on expertise"
    }
  ]
}
\end{Verbatim}

\hrule
\bigskip

\textbf{User Prompt} 

\begin{Verbatim}[breaklines, fontsize=\fontsize{5}{6}\selectfont]
**User action / topic (Context):**
{interpretation}

**Available agents:**
{agents_list}

**Instructions:**
Based on the rules above, select 2 to 4 agents and provide a short reason for each. 
Reply with JSON only, following the specified Output Format.
\end{Verbatim}

\end{tcolorbox}
\caption{Prompt for selecting a subset of relevant agents.}
\label{fig:agent-selection-prompt}
\end{figure*}
\begin{figure*}[h]
\begin{tcolorbox}[width=\linewidth, fontupper=\scriptsize]

\textbf{Discussion Necessity Scoring and Mutter Generation} \\
\hrule
\bigskip

\textbf{System Prompt}
\begin{Verbatim}[breaklines, fontsize=\fontsize{5}{6}\selectfont]
### Task: Score the necessity of a multi-agent discussion (0-1 scale) and optionally produce a "mutter."
For a given workspace change, evaluate the need for three discussion modes: IDEA, DISCUSSION, and EVALUATION. Additionally, generate a brief, personality-driven reaction (mutter).

### Scoring Criteria
- **IDEA**: Need for generating ideas and alternatives.
- **DISCUSSION**: Need for sharing perspectives and addressing conflicts.
- **EVALUATION**: Need for evaluating pros, cons, and criteria.
- **Scale**: Use 0.0-1.0 conservatively. Lower scores (<= 0.2) if a similar discussion already exists nearby.

### Mutter Rules
- **Length**: Maximum 15 characters.
- **Language**: Korean only.
- **Tone**: Must reflect the agent's distinct persona and expertise. Avoid generic phrases like "Great!" or "Interesting."
- **Example**: A finance expert might say "Budget needs review".

### Output Requirements
Output strictly a single JSON object. No markdown, no extra text.
{
  "IDEA": 0.5,
  "DISCUSSION": 0.7,
  "EVALUATION": 0.3,
  "mutter": "Persona-driven phrase"
}
\end{Verbatim}

\hrule
\bigskip

\textbf{User Prompt} 

\begin{Verbatim}[breaklines, fontsize=\fontsize{5}{6}\selectfont]
**Task context (Overall scenario):** {task_description}
**User action / workspace change:** {interpretation}
**Your profile (Agent persona):** Name: {agent_name} / Role: {agent_prompt}
**Existing nearby dialogs**: {existing_nearby_dialogs}

---
Based on the context and your profile, score the discussion necessity and provide a mutter. 
Reply with a single JSON object (IDEA, DISCUSSION, EVALUATION, and optionally mutter).
\end{Verbatim}

\end{tcolorbox}
\caption{Prompt for discussion necessity scoring and mutter generation.}
\label{fig:necessity-scoring-prompt}
\end{figure*}

\end{document}